\documentclass[mcs]{scsthesis}
\tablespagefalse

\usepackage[utf8]{inputenc}
\usepackage{fancyhdr}
\usepackage{caption}
\usepackage{longtable}
\usepackage{amsmath, amsthm, amssymb}
\usepackage[linesnumbered,ruled]{algorithm2e}
\usepackage{graphicx,rotating}
\title{Improved Methods For Generating Quasi-Gray Codes}
\author{Dana Jansens}
\date{}

\newtheorem{thm}{Theorem}[chapter]
\newtheorem{corollary}[thm]{Corollary}
\newtheorem{lemma}[thm]{Lemma}

\everymath{\displaystyle}

\begin{document}
\beforepreface

\prefacesection{Abstract}
Consider a sequence of bit strings of length $d$, such that each string
differs from the next in a constant number of bits.
We call this sequence a quasi-Gray code.
We examine the problem of efficiently generating such codes, by considering
the number of bits read and written at each generating step, the average number
of bits read while generating the entire code, and the number of strings
generated in the code.  Our results give a trade-off between these
constraints, and present algorithms that do less work on average than previous
results, and that increase the number of bit strings generated.

\prefacesection{Acknowledgements}
I must say a big thank you to my supervisors Anil Maheshwari and Michiel Smid,
each of whom must have read parts of this thesis at least ten times by now.
Their support and input has helped a great deal in my success in this degree
program.  I also want to thank Pat Morin, who worked with me on many parts
of this thesis and provided a lot of feedback on the small
details of my writing, even though he was not supervising me and does not
get official credit for it.  I wish to thank Prosenjit Bose and Paz Carmi
as well for input on these problems, and Lucia Moura and Brett Stevens for
being on my thesis defence committee.  I must also send a thank you to Anil
Somayaji, who invited me to join in writing a paper during my third year
of undergrad studies.  I never anticipated at the time that it would
lead to me attending grad school and eventually writing an entire thesis.
My inspiration and ability to write this thesis can be attributed
in no small way to role models such as these.

I want to give a big shout-out to the entire
Carleton computational geometry lab crew,
with who I eat lunch every day, who listen to my research presentations,
discuss research ideas with me,
and who make me feel welcome and a legitimate part of the group.
Their presence has been vital to me, and I know I would have struggled
much more in this degree without them.  I especially want to thank
Prosenjit Bose as he goes out of his way to keep the group together and
to make the space welcoming to us all.

I have been fortunate to find a lot of support and encouragement in my life,
without which I know this accomplishment would not be possible.
Djamila Ibrahim, Heather Larke, Larry and Sue Larke, and Lisa Cooper have
all been there to support me, believe in me, push me, and catch me, and I am
very grateful for it all.

I will continue to push myself on to my next, bigger, challenge thanks
to all of these people.
\afterpreface

\chapter{Introduction}
\label{sec:intro}

\section{Problem Statement}
\label{subsec:problem_statement}

We are interested in efficiently generating a sequence of bit
strings.  The class of bit strings we wish to generate
are cyclic quasi-Gray codes.  A
\emph{Gray code} \cite{Gray53} is a sequence of bit strings,
such that any two consecutive strings differ in exactly one bit.
We use the term \emph{quasi-Gray code} \cite{Fredman78} to refer
to a sequence of
bit strings where any two consecutive strings differ in at most $c$ bits,
where $c$ is a constant defined for the code.  A Gray code (quasi-Gray code)
is called \textit{cyclic} if the first and last
generated bit strings also differ in at most $1$ bit ($c$ bits).

We say a bit string that contains $d$ bits has \textit{dimension} $d$, and are
interested in efficient algorithms to generate a sequence of bit strings
that form a quasi-Gray code
of dimension $d$.
After generating a bit string,
we say the algorithm's data structure corresponds exactly to the
generated bit string, and it's \textit{state} is the bit string itself.
In this way, we restrict an algorithm's data structure to using exactly
$d$ bits.
At each step, the input to the algorithm
will be a bit string, which is the algorithm's
current state.
The output will be a new bit string that corresponds to the next state of the
algorithm's data structure.


The number of consecutive unique bit strings generated is equal to the
number of consecutive unique states for the generating data structure,
and we call this value $L$, the \textit{length} of
the generated code.  Clearly $L \le 2^d$.
We define the \emph{space efficiency} of an algorithm as the ratio $L/2^d$,
that is, the fraction of bit strings generated out of all possible
bit strings given the dimension of the strings.
When the space efficiency is $1$, we call the
data structure \emph{space-optimal}, as it generates all possible bit strings.
When $L < 2^d$, the
structure is non-space-optimal; as we will see, this allows the time required
to generate each consecutive bit string to be improved.

Each generating step takes as input the output of the previous generating step,
which is a bit string in the quasi-Gray code.  The average number of bits
read is defined to be the ratio of the
total number of bits read, to the length of the code, when generating
one iteration of the entire quasi-Gray code.

Our goal is to study and improve efficiency of algorithms for
generating quasi-Gray codes in the following ways.
\begin{enumerate}
 \item Worst-Case read:
We would like to know how many bits the algorithm must read in the
worst case in order to make the appropriate changes in the input string and
generate the next bit string in the code, and find ways to reduce this
when possible.
 \item Worst-case write:
We would like to know how many bits must change in the worst case
to reach the successor string in the code, and keep this to $1$ when possible
while improving upon other metrics.
 \item Average number of bits read:
We would like to reduce the average number of bits read
at each generating step while maintaining other metrics.
 \item Space efficiency:
We would like our algorithms to
be as space efficient as possible, ideally generating as many bit strings
as their dimension allows, with $L=2^d$.
\end{enumerate}
Our results give a trade-off between
these different goals.

Our decision to limit the algorithm's data structure to exactly $d$ bits
differs from previous work, where the data structure could use
more bits than the strings
it generated \cite{Fredman78, Rahman08}.  To compare previous results to
our own, we consider the extra bits in their data structure to be a part of
their generated bit strings.
This gives a more precise view of the space efficiency of an algorithm.

Each generated bit string of dimension $d$ has a distinct totally
ordered rank in the generated code with respect to the initial bit string in
the code.  For a cyclic code, the initial bit string can be chosen
arbitrarily.  We assume the initial bit string to be
the bit string of $d$ zeros unless stated otherwise.
Given a string of rank $k$ in a code of length $L$, where $0 \le k < L$,
we want to support the following operations:
\begin{itemize}
	\item \emph{next} generates the bit string of rank $(k + 1) \mod L$
	\item \emph{previous} generates the bit string of rank $(k - 1) \mod L$
\end{itemize}

We work within the bit probe model \cite{perceptrons, Rahman08},
where the performance of an algorithm is measured by counting the
average-case and the worst-case number of
bits read and written.  We examine these values for the process of generating
each bit string in a quasi-Gray code.
We use the Decision Assignment
Tree (DAT) model \cite{Fredman78} (which we describe further in
Chapter \ref{sec:dats}) to construct algorithms for
generating quasi-Gray
codes and describe the algorithms' behaviour, as well as to
discuss upper and lower bounds.

\section{Definitions}
\label{subsec:defns}
We use a notation for the iterated logarithm function of the
form $\log^{(c)}n$ where $c$ is a non-negative integer, and is always
surrounded by brackets to differentiate it from an exponent.  The value of
the function is defined as follows.  When $c=0$, $\log^{(c)}n = n$.  If
$c>0$, then $\log^{(c)}(n) = \log^{(c-1)}(\log(n))$.  For example,
$\log^{(2)}n = \log(\log n)$.  Throughout, the base of the $\log$
function is assumed to be $2$ unless stated otherwise.

We define the function $\log^*n$ to be equal to the smallest non-negative value
of $c$ such that $\log^{(c)}n \le 1$.  For example $\log^*1 = 0$ and
$\log^*3 = 2$.

\section{Results Summary}
\label{subsec:resultssum}

\begin{sidewaystable}
\begin{center}
\begin{tabular}[]{|c|c|c|c|c|c|} \hline
	\rule{0pt}{3ex} \rule[-1.2ex]{0pt}{0pt}
	&& \multicolumn{2}{|c}{Bits Read} & \multicolumn{1}{|c}{Bits Written} & \multicolumn{1}{|c|}{} \\ \cline{3-5}
	\multicolumn{1}{|c}{ \rule{0pt}{3ex} \rule[-1.2ex]{0pt}{0pt} Dimension} &
	\multicolumn{1}{|c}{Space Efficiency} &
	\multicolumn{1}{|c}{Average} &
	\multicolumn{1}{|c}{Worst-Case} &
	\multicolumn{1}{|c}{Worst-Case} &
	\multicolumn{1}{|c|}{Reference} \\
	\hline

	\rule{0pt}{3ex} \rule[-1.2ex]{0pt}{0pt}
	$d$ & 
	$1$ & 
	$2 - 2^{1-d}$ & 
	$d$ & 
	$d$ & 
	folklore \\
	\hline

	\rule{0pt}{3ex} \rule[-1.2ex]{0pt}{0pt}
	$d$ & 
	$1$ & 
	$d$ & 
	$d$ & 
	$1$ & 
	\cite{Fredman78, Gray53} \\
	\hline

	\rule{0pt}{3ex} \rule[-1.2ex]{0pt}{0pt}
	$d$ & 
	$1$ & 
	$6\log d$ & 
	$d$ & 
	$1$ & 
	Theorem \ref{thm:partcode} \\
	\hline

	\rule{0pt}{3ex} \rule[-1.2ex]{0pt}{0pt}
	$d$ & 
	$1$ & 
	$6\log^{(2c-1)} d+11$ & 
	$d$ & 
	$c$ & 
	Theorem \ref{thm:supercode} \\
	\hline

	\rule{0pt}{3ex} \rule[-1.2ex]{0pt}{0pt}
	$d$ & 
	$1$ & 
	$17$ & 
	$d$ & 
	$\lfloor(\log^*d+5)/2\rfloor$ & 
	Corollary \ref{cor:logstar} \\
	\hline

	\rule{0pt}{3ex} \rule[-1.2ex]{0pt}{0pt}
	$d$ & 
	$1/2$ & 
	$O(1)$ & 
	$\log d + 4$ & 
	$4$ & 
	\cite{Rahman08} \\
	\hline

	\rule{0pt}{3ex} \rule[-1.2ex]{0pt}{0pt}
	$d=n+\log n$ & 
	$2/n - O(2^{-n+1})$ & 
	$3$ & 
	$\log n + 1$ & 
	$\log n + 1$ & 
	\cite{256309} \\
	\hline

	\rule{0pt}{3ex} \rule[-1.2ex]{0pt}{0pt}
	$d=n+\log n$ & 
	$1/2 + O(1/2n)$ & 
	$4$ & 
	$\log n + 1$ & 
	$\log n + 1$ & 
	\cite{Brodal} \\
	\hline

	\rule{0pt}{3ex} \rule[-1.2ex]{0pt}{0pt}
	$d=n+(t+1)\log n$ & 
	$1-O(n^{-t})$ & 
	$O(1)$ & 
	$(t+1)\log n + 1$ & 
	$(t+1)\log n + 1$ & 
	Corollary \ref{cor:doublespint} \\
	\hline

	\rule{0pt}{3ex} \rule[-1.2ex]{0pt}{0pt}
	$d=n+(t+1)\log n$ & 
	$1-O(n^{-t})$ & 
	$O(t\log n)$ & 
	$(t+1)\log n+1$ & 
	$3$ & 
	Corollary \ref{cor:winebrgct} \\
	\hline

	\rule{0pt}{3ex} \rule[-1.2ex]{0pt}{0pt}
	$d=n+(t+1)\log n$ & 
	$1-O(n^{-t})$ & 
	$12\log^{(2c)}n + O(1)$ & 
	$(t+1)\log n + 1$ & 
	$2c+1$ & 
	Corollary \ref{cor:winet} \\
	\hline

	\rule{0pt}{3ex} \rule[-1.2ex]{0pt}{0pt}
	$d=n+(t+1)\log n$ & 
	$1-O(n^{-t})$ & 
	$O(1)$ & 
	$(t+1)\log n + 1$ & 
	$\log^*n-3$ & 
	Corollary \ref{cor:winelogstar} \\
	\hline

\end{tabular}

 \caption{
	Summary of results.
	When ``Worst-Case Bits Written'' is a constant then the resulting code
	is a quasi-Gray code, and when it is $1$, the code is a Gray code.
	$c \in \mathbb{Z}$
	and $t$ are constants greater than $0$.}
 \label{table:results}
\end{center}
\end{sidewaystable}

Our results, as well as previous results,
are summarized in Table \ref{table:results}.

First, we present some space-optimal algorithms.  Although our space-optimal
algorithms read a small number of bits in the average case, they all read
$d$ bits in the worst case.

In Section \ref{subsec:partition}, we describe the
Recursive Partition Gray Code (RPGC) algorithm,
which generates a Gray code of dimension $d$ while reading on
average no more than $6\log d$ bits.
This improves the average number of bits read for a space-optimal Gray code
from $d$ to $O(\log d)$.
In Section \ref{subsec:composite}, we use the RPGC
to construct a DAT that generates a quasi-Gray code
while reducing the average number of bits read.  We then
apply this technique iteratively in Section \ref{subsec:rpgc-composite} to
create, for any constant $c \ge 1$, a $d$-dimensional DAT that reads
worst-case $d$ bits, but reads
on average only $6 \log^{(2c-1)}d + 11$ bits, and writes at most $c$ bits.
This lowers the average number of bits read
to generate
a space-optimal quasi-Gray code from $O(\log d)$ to $O(\log^{(2c-1)} d)$,
when $c \ge 2$.

In Section \ref{subsec:const-read} we create
a $d$-dimensional DAT that reads worst-case all $d$ bits, while
reading at most $17$ bits on average, and writing at most
$\lfloor (\log^*d+5)/2 \rfloor$ bits
to generate each bit string, while also being space-optimal.  This reduces the
average number of bits read to $O(1)$ for a space-optimal code,
but increases the number of bits written to be slightly more than a constant.

Next, we consider quasi-Gray codes that are not space-optimal, but achieve
space efficiency arbitrarily close to $1$, and that read $O(\log d)$ bits in the
worst case.

In Section \ref{subsec:lazycounters} we construct a
DAT of dimension $d=n+(t+1)\log n$
that reads and writes $(t+1)\log n+1$ bits
in the worst case, and has space efficiency $1-O(n^{-t})$.
This improves the space efficiency dramatically of previous results,
when the worst-case number of bits written is $O(\log n)$,
from $1/2$ to $1$ in the limit $n \to \infty$.
By combining a simple Gray code with this result, we are able to produce a DAT
of dimension $d=n + (t+1)\log n$ that reads $O(t\log n)$ bits on
average and in the worst case, but writes at most $3$ bits.  This reduces
the worst-case number of bits written from $O(\log n)$ to $O(1)$,
while the space efficiency remains asymptotically the same.

We then combine results
from Section \ref{subsec:rpgc-composite} to produce
a DAT of dimension $d=n+(t+1)\log n$
that reads in the worst case $(t+1)\log n + 1$ bits, reads on
average $12\log^{(2c)}n + O(1)$ bits, and writes at most $2c+1$ bits, for
any constant $c \ge 1$.  This improves the average number of
bits read to generate
a quasi-Gray code with $O(\log d)$ bits read in the worst case.  We
reduce the average number of bits read from
$O(\log d)$ to $O(\log\log d)$ when writing the same number of bits,
and when writing a constant number more, the average becomes $O(\log^{(2c)}d)$,
while keeping the space efficiency arbitrarily close to $1$.
Lastly, we show this DAT can also generate a code while reading a constant
number of bits on average, if it writes $\log*n-3$ bits in the worst case.

A summary of the results in this thesis is appearing at SWAT 2010, the
12th Scandinavian Symposium and Workshops on Algorithm Theory.

\section{Organization of the thesis}
\label{subsec:organization}

The remainder of this work will be organized as follows.
In Chapter \ref{sec:previous_work}, we review
related previous work, and discuss its relationship to our own work.
In Chapter \ref{sec:dats}, we present the Decision Assignment Tree model and
make some observations about generating quasi-Gray codes within the model.
In Chapter \ref{sec:results}, we present our results, starting with the
Recursive Partition Gray Code, followed by the RPGC-Composite quasi-Gray Code,
and finally our Lazy Counters which build up to
our \texttt{WineIncrement} counter.
We conclude in Chapter \ref{sec:conclusion}, with a summary of our work
and discussion of related open problems and future work.

\chapter{Previous work}
\label{sec:previous_work}

\section{Gray codes}
\label{subsec:gray_codes}

The Gray code was invented by Frank Gray in 1953 \cite{Gray53}.  In his
patent application, Gray described what we now know as the Binary Reflected
Gray Code (BRGC).  This code is a sequence of bit strings of dimension $d$,
where each successive bit string differs from the previous one in exactly one
bit and where all the possible bit strings are present.
From this code came the more general term, Gray code, which refers to any
sequence of bit strings where successive strings differ in exactly one bit.
Furthermore, the concept of
a cyclic Gray code was used to describe a Gray code where
the first and last bit strings differ in exactly one bit, such as the
original BRGC, creating a secondary class of non-cyclic Gray codes.

The BRGC has a structure which can be defined recursively.
For a single bit, the code is simply $0$ followed by $1$.  To
create a code of $d+1$ bits, given the code of $d$ bits: first place
the $2^d$ bit strings of the dimension $d$ in order.  Concatenate a $0$ onto
the left end of each of the bit strings.  Then repeat the same $2^d$ bit strings
in reverse order, concatenating a $1$ onto the left end of each one.  Figure
\ref{fig:binarygraycode} shows the BRGC for up to three bits.  Note that the
last bit string in the code differs from the first in a single bit, making the
code a cyclic Gray code.

\begin{figure}
	\begin{center}
	\begin{tabular}{p{3cm}p{3cm}p{3cm}}
		\begin{center}
		\begin{tabular}{c}
			0\\
			\hline
			1
		\end{tabular}
		\end{center}
		&
		\begin{center}
		\begin{tabular}{cc}
			0 & 0 \\
			0 & 1 \\
			\hline
			1 & 1 \\
			1 & 0
		\end{tabular}
		\end{center}
		&
		\begin{center}
		\begin{tabular}{ccc}
			0 & 0 & 0 \\
			0 & 0 & 1 \\
			0 & 1 & 1 \\
			0 & 1 & 0 \\
			\hline
			1 & 1 & 0 \\
			1 & 1 & 1 \\
			1 & 0 & 1 \\
			1 & 0 & 0
		\end{tabular}
		\end{center}
	\end{tabular}
	\end{center}
	\caption{The standard Binary Reflected Gray Code for 1, 2, and 3 dimensions}
	\label{fig:binarygraycode}
\end{figure}

\section{Quasi-Gray codes}
\label{subsec:quasi_gray_codes}

A further generalization of Gray codes was provided by Fredman \cite{Fredman78}
when he coined the term \emph{quasi-Gray code}.  In Fredman's work, a quasi-Gray
code was defined to be a $d$-dimensional bit string along with some unbounded
additional data structure.  The quasi-Gray code differs in each
successive state in one bit of the $d$-dimensional bit string,
but the algorithm may also make arbitrary changes to its additional
data structure.
In order to efficiently generate the successor bit string in a quasi-Gray code,
the algorithm may make use of the additional data structure, reading fewer bits
than it would otherwise need to.  Fredman characterizes the efficiency of
a generating algorithm by the worst-case number of bits read and written to
generate each bit string in the code.  Under this model, the algorithm will
write a single bit in the worst case if and only if the algorithm does not
use any additional data structure.  In this case, Fredman notes that
it would be generating a Gray code.

We modify this definition slightly to improve the clarity of analysis.
Because the additional data
structure and the code itself are equally part of the algorithm's state, we
consider Fredman's additional data structure to be a part of the generated
code.
We use the term quasi-Gray code to refer to a sequence of $d$-dimensional
bit strings where successive bit strings each differ by at most a constant
number of bits.
This is similar
to Fredman's model, but with a stronger limitation on the number of bits being
written, as his model allowed an arbitrary number of bits to be written while
still considering it a quasi-Gray code.
Thus an algorithm that, under Fredman's definition,
generated a $d$-dimensional quasi-Gray code using
an additional $k$ bits of data structure and writing at most a constant number
of bits would, in our redefinition of the term,
simply generate a $(d+k)$-dimensional quasi-Gray code.

Fredman's results generate codes while writing in the worst case $O(2^d)$ bits,
which do not qualify as quasi-Gray codes under our definition of the term, and
which have space efficiency $O(1/2^{2d})$.

Rahman and Munro \cite{Rahman08} begin to address space efficiency while
generating quasi-Gray codes, without naming it explicitly, which we continue
in this thesis.  However, we expand upon their work by also
examining the average number of bits read by generating algorithms.
Rahman and Munro construct quasi-Gray codes in the manner of Fredman, where
they generate a $d$-dimensional Gray code, while keeping some additional
data structure.  However, the authors use much smaller data structures,
eventually constructing an algorithm that requires only three extra bits.
This brings the space efficiency of their algorithm up to $1/8$, which was
the first algorithm to read less than $d$ bits in the worst case with
space efficiency that did not become arbitrarily close to $0$ for large $d$.
This algorithm generates a $d$-dimensional quasi-Gray code
while reading $\log (d-3)+6$ bits in the worst case
and writing at most $7$ bits.

Rahman and Munro do not use the DAT model, and they sometimes include in their
analysis
the amount of work to decode rankings inside the quasi-Gray code's structure.
Under the DAT model, the rank can be embedded
into the tree, so we don't need to examine such costs.  For this reason
we compare our work to theirs only in the case where they also ignore these
costs.  Rahman and Munro give one such algorithm, which ignores the work
involved in determining ranks.  The bounds given for this algorithm also
hold in the DAT model.  The algorithm uses only a single bit of extra
data structure, giving it a space efficiency of $1/2$.
This algorithm generates a $d$-dimensional quasi-Gray code while
reading $\log (d-1)+4$ bits in the worst case
and writing at most $4$ bits.

Rahman and Munro also consider the problem of adding or subtracting two numbers,
when each is stored using a quasi-Gray code representation.
They give a data structure that uses $d = n + O(\log^2 n)$ bits.  Incrementing
or decrementing a number on its own, equivalent to
generating the next or previous
bit string in the quasi-Gray code, requires at most $O(\log n)$ bits to be read
and at most $5$ bits to be changed.  This data structure also supports
adding or subtracting two such numbers of dimension
$d$ and $d'$, where $d \ge d'$, while reading at most $O(d + \log d')$ bits.

Savage highlights works around Gray codes in her
survey \cite{Savage}, which includes both Gray codes as sequences
of bit strings, and more general combinatorial sequences with minimal
change between successive states, referred to as \emph{combinatorial Gray codes}.
The majority of the work related to
quasi-Gray codes for bit strings discusses
mathematical aspects of Gray codes such as existence of various classes of
Gray codes, rather than algorithms for efficient generation of these sequences.

Frank Ruskey devotes a chapter of his book-in-progress \cite{RuskeyWIP} to
algorithms for generating combinatorial Gray codes.  The majority of this
work is devoted to generating other forms of combinatorial Gray codes than
bit strings as we consider in this thesis, but he does include a section on
generating the BRGC.  In this section, Ruskey describes an algorithm to generate
the $d$-dimensional BRGC while reading and writing $O(1)$ bits
in the worst case, by making use of an additional $d$ bits.
This gives a quasi-Gray code with space efficiency of $1/2^d$.

Knuth, in volume $4$ of The Art of Computer Programming \cite{Knuth},
discusses the problem of generating Gray codes.  He gives an overview of
various applications for Gray codes, and surveys some known results,
both in generating them, and in analysis of other properties.  Knuth shows
an algorithm for \emph{loopless} generation of a Gray code of
dimension $d$, \cite{DBLP:journals/cacm/BitnerER76}
where each generating step can be executed without any loop in the algorithm,
that has space efficiency $2^{-d}$.  He also discusses other properties of
Gray codes, such as a balanced number of bit flips for each bit position,
having each bit keep its value for at least a constant number of states,
or \emph{monotonicity}, where each bit string of rank $x$ in the code has
at most as many bits set to $1$ as the bit string of rank $x+2$.

Frandsen \textit{et al.} \cite{256309} describe a method of generating
a sequence of bit strings of dimension $d=n+\log n$
while reading and writing at most $O(\log n)$ bits for each generating step.
Their counter algorithm
has space efficiency $O(1/n)$, which converges to $0$ in the
limit $n \to \infty$.
An observation by Brodal \cite{Brodal} improves the space efficiency
of their counter algorithm to $1/2$, matching the efficiency of the work by
Rahman and Munro.
We improve on these counters in Section \ref{subsec:lazycounters} in terms of
average bits read and space efficiency.

\section{Upper and lower bounds}
\label{subsec:bounds}

We use the Decision Assignment Tree (DAT) model to analyze algorithms
for generating bit strings in a quasi-Gray code.  The model was first introduced
for this context by Fredman \cite{Fredman78}.
We will describe the DAT model and how we use it in detail in
Chapter \ref{sec:dats}, while briefly describing it here.

An algorithm to generate a quasi-Gray code takes a bit string of
dimension $d$ as input, and modifies it to
become the next bit string in the quasi Gray-code.  This
operation does not necessarily require reading all $d$ bits of the current
string.
The DAT model can be used for proving both upper and lower bounds
on the required number of bits to be read in order to generate a quasi-Gray code.
In this work, we construct and analyze our algorithms under this model to
provide upper bounds.

Meanwhile, a non-trivial lower bound for the worst case number of bits read
while generating a Gray code remains unknown.
A trivial DAT, such as iterating through the standard binary
representations of $0$ to $2^d-1$, in the worst case, will require reading and
writing all $d$ bits to generate the next bit string, but it may also read
and write as few as one bit when the least-significant bit changes.
On average, it reads and writes $2-2^{1-d}$ bits.
Meanwhile, it is possible to create a DAT that
generates the Binary Reflected Gray Code, as described in
Section \ref{subsec:brgcdat}.  This DAT would always write exactly
one bit, but requires reading all $d$ bits to generate each successive bit
string in the code.
This is because the least-significant bit is flipped if and only if
the parity is even, which can only be determined by reading all $d$ bits.

To generate a Gray code of dimension $d$ with length $L=2^d$,
Fredman \cite{Fredman78} uses the DAT model to show
that any algorithm will require reading $\Omega(\log d)$ bits for
some bit string.
Fredman conjectures
that for a Gray code of dimension $d$ with $L=2^d$, any DAT will have to
read all $d$ bits to generate at least one bit string in the code.
That is, any DAT
generating the code must have height $d$.  This remains an open
problem.\footnote{In \cite{Rahman08} the authors claim to have proven this
conjecture true for ``small'' $d$ by exhaustive search.}

\chapter{Decision Assignment Trees}
\label{sec:dats}

\section{The Decision Assignment Tree Model}
\label{subsec:the_dat_model}

\begin{figure}
\label{fig:brgcdat}
\begin{center}
\includegraphics[width=\linewidth]{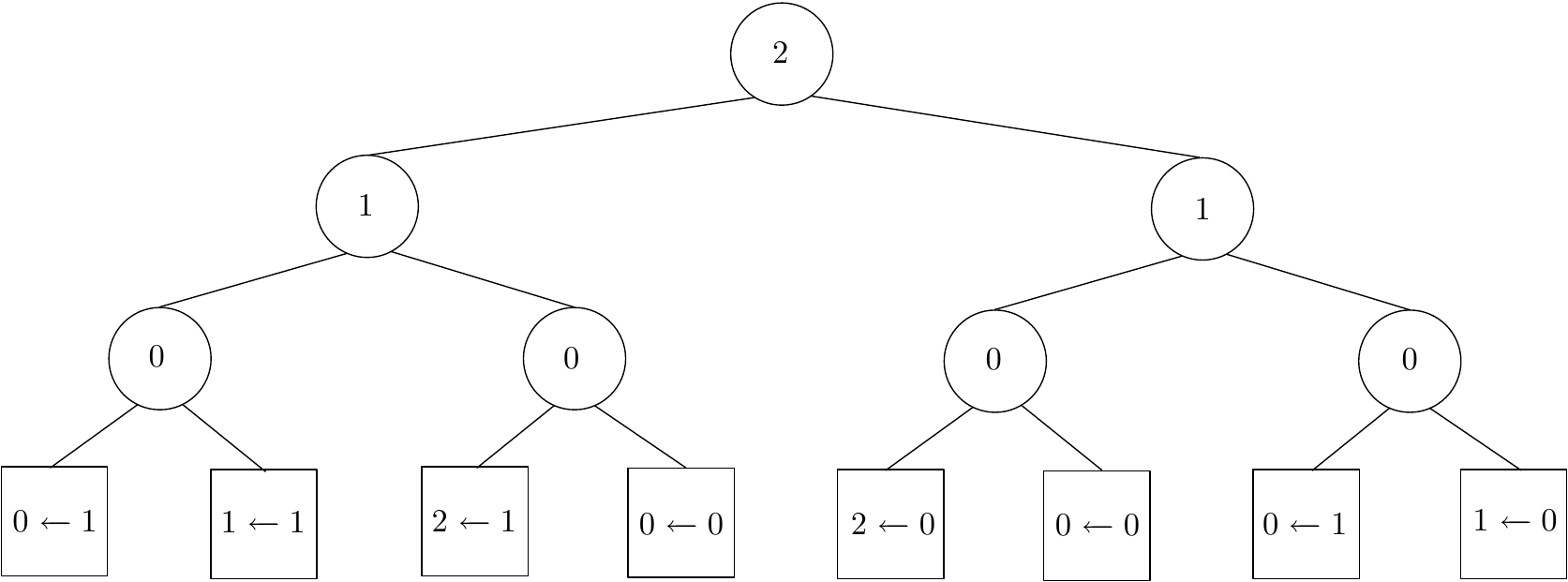}
\end{center}
\caption{A DAT that generates the Binary Reflected Gray Code on three bits.
The bits are labelled from $0$ as the right-most to $2$ as the left-most bit.
The label in each internal node specifies which bit is being read.  Control
moves to the left child or right child if the bit is equal to $0$ or $1$,
respectively.
The rules in the leaf nodes set a bit $i$ to a new value, $0$ or $1$. }
\end{figure}

In the DAT model, an algorithm is
described as a binary tree.  We say that a DAT which reads and
generates bit strings of length $d$ has \emph{dimension} $d$.  Further, we
refer to the bit string that the DAT reads and modifies as
the \emph{state} of the DAT.  Generally the initial bit string for
a quasi-Gray code of dimension $d$, and thus the initial state of
its generating DAT, is the bit string made up of a sequence of $d$ zeros.
Each internal node of the tree
is labeled with a single fixed position $0 \le i \le d-1$
within the input bit string, and represents reading that bit $i$.
Figure \ref{fig:brgcdat} shows a DAT that generates the BRGC of
dimension $d=3$.  The BRGC that is generated is also seen in
Figure \ref{fig:binarygraycode}.

Let $T$ be a DAT of dimension $d$.  The algorithm starts at the
root of $T$, and reads the bit with which that node is labeled.  Then it
moves to a left or right child of that node, depending on whether
the bit read was
a $0$ or a $1$, respectively.
This repeats recursively until a leaf node in the tree is
reached.

Each leaf node of $T$ represents a subset of states
where the bits read along the path to the leaf are in a fixed
state.
More formally, a state $S_i$
for a DAT is represented by a single leaf $L_i$.
When the DAT is in state $S_i$, traversing the DAT while reading the
current state will lead to the leaf $L_i$.  It is possible for
two different states $S_i$ and $S_j$ to share the same leaf $L_m$ if
they both cause the same bit
positions to be read and those positions share all the same values.  In
this case it is required at least one of
the bits that were not read on the path to
the leaf $L_m$ must be in a different state each time the DAT traversal
reaches $L_m$.

The leaf nodes each contain rules that describe which bits to update to
generate the next bit string in the code.
The update rules are constrained in the following ways:
\begin{enumerate}
	\item Each rule must set a single fixed bit directly to $0$ or to $1$.
	\item The rules together must change at least one bit.
\end{enumerate}

Analysis under this model can be done by examining the structure of the tree.
The worst-case
number of bits read will be equal to the height of the tree, and the
worst-case number of bits written will be equal to the maximum number of rules
in any leaf of the tree.  The average number of bits read and written are
not easily derived from the tree's structure.  The average number of bits
read will be equal to the sum of all paths in the tree, weighted by the
fraction of times the path is used when generating the entire quasi-Gray code.
The average number of bits written will be a similar weighted average,
for the number of rules in the leaves.

\section{Assembling DATs}
\label{subsec:assemblingdats}

Decision Assignment Trees can be assembled by joining
together other DATs.  We present here observations based on this.

\begin{lemma}\label{lem:recursivedattree}
	Let $L$ and $R$ each be a DAT for a binary code of dimension $d$ with
space efficiency $1$.
The $L$ and $R$ trees may be joined together, under a new
root node, to create a DAT of dimension $d+1$ with space efficiency $1$.
\end{lemma}

\begin{proof}\label{pf:recursivedattree}
	We join the $L$ tree and $R$ tree together by adding a new root
node, and making $L$ and $R$ its left and right subtrees respectively.  The
subtrees $L$ and $R$ each read and write $d$ bits.  We assign them to the
same bits, $0$ to $d-1$, and the root node to the $d$\emph{-th} bit.  We
assume w.l.o.g. that reading a $0$ at the root node means to move to the
root of the $L$ subtree, while $1$ means to move to the root of the $R$
subtree.  Assume that the $d$\emph{-th} bit is initially set to $0$.  If it
is $1$, then swap $L$ and $R$ in what follows.
	
	It is clear that $L$ remains a valid Decision Assignment Tree of
dimension $d$.  And because it never changes the $d$\emph{-th} bit,
it will never cause the $R$ subtree to be used.  Thus our initial
construction is a valid Decision Assignment Tree of dimension $d+1$ that
counts through only $2^d$ states.
	
	Let $L_A$ be the first state of $L$ and $L_Z$ be the last state. 
If $L$ is cyclic, then any two states such that $L_A$ immediately follows
$L_Z$ in the code are valid.  Similarly, let $R_A$ and $R_Z$ be the
first and last states of $R$.  We modify the construction to
join the code generated by $L$ to the code generated by $R$, producing
a new code of dimension $d+1$:
\begin{enumerate}
	\item Make the update rules of $L_Z$ change the counter to state
$R_A$, and change the $d$\emph{-th} bit to $1$.
	\item Make the update rules of $R_Z$ change the counter to state
$L_A$, and change the $d$\emph{-th} bit to $0$.
\end{enumerate}

	Note that the leaf for state $L_Z$ may be shared by another state $L_x$,
and is therefore invoked multiple times by the $L$ subtree.
If this is the case, simply split the leaf node, giving it
two children that differentiate on a bit that is different in $L_Z$ and $L_x$.
Repeat this process until the leaf for $L_Z$ is not used for any other states
in $L$.

	Because the $L$ subtree is able to count through $2^d$ states, it
will take $2^d-1$ steps to go from $L_A$ to $L_Z$, and likewise
for the $R$ subtree.  Within each of these steps, the
$d$\emph{-th} bit is not changed and the subtree is able to operate
correctly.  After generating $2^d-1$ bit strings in $L$, the $d$\emph{-th}
bit is changed, and the state is changed to $R_A$.  This makes $2^d$
consecutive bit strings, generated by the $L$ subtree.  Now the
same argument holds for the $R$ subtree, which will generate $2^d-1$
bit strings within its own subtree, and then move to $L_A$,
completing a full cycle through all $2^{d+1}$ possible bit strings.
Thus, we have a cyclic binary code of dimension $d+1$ and space efficiency $1$.
\end{proof}

\section{Generating the BRGC}
\label{subsec:brgcdat}

\begin{lemma}\label{lem:graycodeasdat}
	The Binary Reflected Gray Code of dimension $d$ can be generated by a DAT,
which requires reading $d$ bits and writing at most $1$ bits
to generate each successive bit string.
\end{lemma}

\begin{proof}\label{pf:graycodeasdat}
The proof is by induction.
For a Binary Reflected Gray Code of dimension $1$, create a DAT
with height $1$.  At the root node, the single bit is read.
If the root reads a $0$, move to its left child, if it reads a $1$, move to its
right child.  The left child changes bit $0$ to $1$ and the right child
changes bit $0$ to be $0$.  This generates the cyclic BRGC of dimension $1$.

Let $L$ be a DAT for the BRGC of dimension $d$, and let $R$ be a DAT which
generates the same bit strings as $L$ in reverse order.
Then, by Lemma \ref{lem:recursivedattree}, we can use $L$ and
$R$ to construct a new DAT of dimension $d+1$.  We choose $L_A$ to be the
bit string $000...0$ and $L_Z$ to be the bit string $1000...0$.
Since the BRGC is
cyclic, $L_A$ is the state which follows $L_Z$.
We choose $R_A$ to be the state $1000...0$
and $R_Z$ to be $000...0$.

Because $R_Z = L_A$ and $L_Z = R_A$, no bits need to be
changed to move between them.  Thus, in the combined DAT, only
the $(d+1)$\emph{-th}
bit needs to change in order to generate $L_A$ from $R_Z$ or $R_A$ from $L_Z$,
and the DAT is able to move between subtrees with only one bit changed.
Further, the first $2^d$ states will correspond to a dimension $d$ BRGC,
with a $0$ in the $(d+1)$\emph{-th} bit.  The second $2^d$ states
will correspond to a dimension $d$ BRGC in reverse order, with a $1$
in the $(d+1)$\emph{-th} bit.  This is precisely the definition of the BRGC
of dimension $d+1$.  Thus, we are able to construct a DAT that generates the
BRGC of any dimension.
\end{proof}

An example of a DAT that generates the BRGC for dimension $d=3$ can be seen
in Figure \ref{fig:brgcdat}.

\chapter{Efficient generation of quasi-Gray codes}
\label{sec:results}

In this chapter we address how to efficiently generate
quasi-Gray codes of dimension $d$.
We examine efficiency in terms of the number of
bits read and written in the worst case to generate each successive bit string,
the number of bits read on average to generate each successive bit string
while generating the entire code, and the space
efficiency.
The codes we generate are all cyclic.  First we present DATs that
read up to $d$ bits in the worst case, but
read fewer bits on average.  Then we present our lazy counters that read
at most $O(\log d)$ bits in the worst case, while also reading
fewer bits on average.

\section{Recursive Partition Gray Code (RPGC)}
\label{subsec:partition}

We show a method for generating a cyclic Gray code of dimension $d$
that requires reading
an average of $6\log d$ bits to generate each successive bit string.
First, assume that $d$ is a power of two for simplicity.  In this special
case, we show that it reads on average no more than $4\log d$ bits to
generate each bit string in the code.
We use both an \emph{increment} and \emph{decrement} operation
to generate the Gray code, where the increment operation generates the next
bit string of dimension $d$ in the code, and the decrement operation generates
the previous bit string of dimension $d$ in the code.

Both increment and decrement operations are defined recursively,
and they
make use of each other.  Pseudocode for these operations is provided in
Algorithms \ref{proc:recur-inc2} and \ref{proc:recur-dec2}.
To generate the next bit string in the code, we partition the bit string of
dimension $n$ into two substrings, $A$ and $B$,
each of dimension $n/2$.  We then recursively increment $A$ unless $A=B$,
that is, unless
the bits in $A$ are in the same state as the bits in $B$, at which point we
recursively decrement $B$.
Testing $A=B$ is done by reading and comparing sequential pairs of bits in $A$
and $B$ until a pair is found that differ.  In the analysis, we will see
that this test reads only a constant number of bits on average.

To generate the previous bit string in the code,
we again partition the bit string of dimension $n$
into two substrings, $A$ and $B$, each of dimension $n/2$.  We then recursively
decrement $A$ unless $A = B+1$, that is, the bits of $A$ are in the same state
as the bits of $B$ would be after an increment operation, at which time we
recursively increment $B$ instead.
Testing $A=B+1$ can be done by simulating an increment of $B$ and
testing $A$ for equality against the result.  In the analysis, we will see
that this test also reads only a constant number of bits on average.

\begin{figure}
 \centering
 \includegraphics[width=0.75\linewidth]{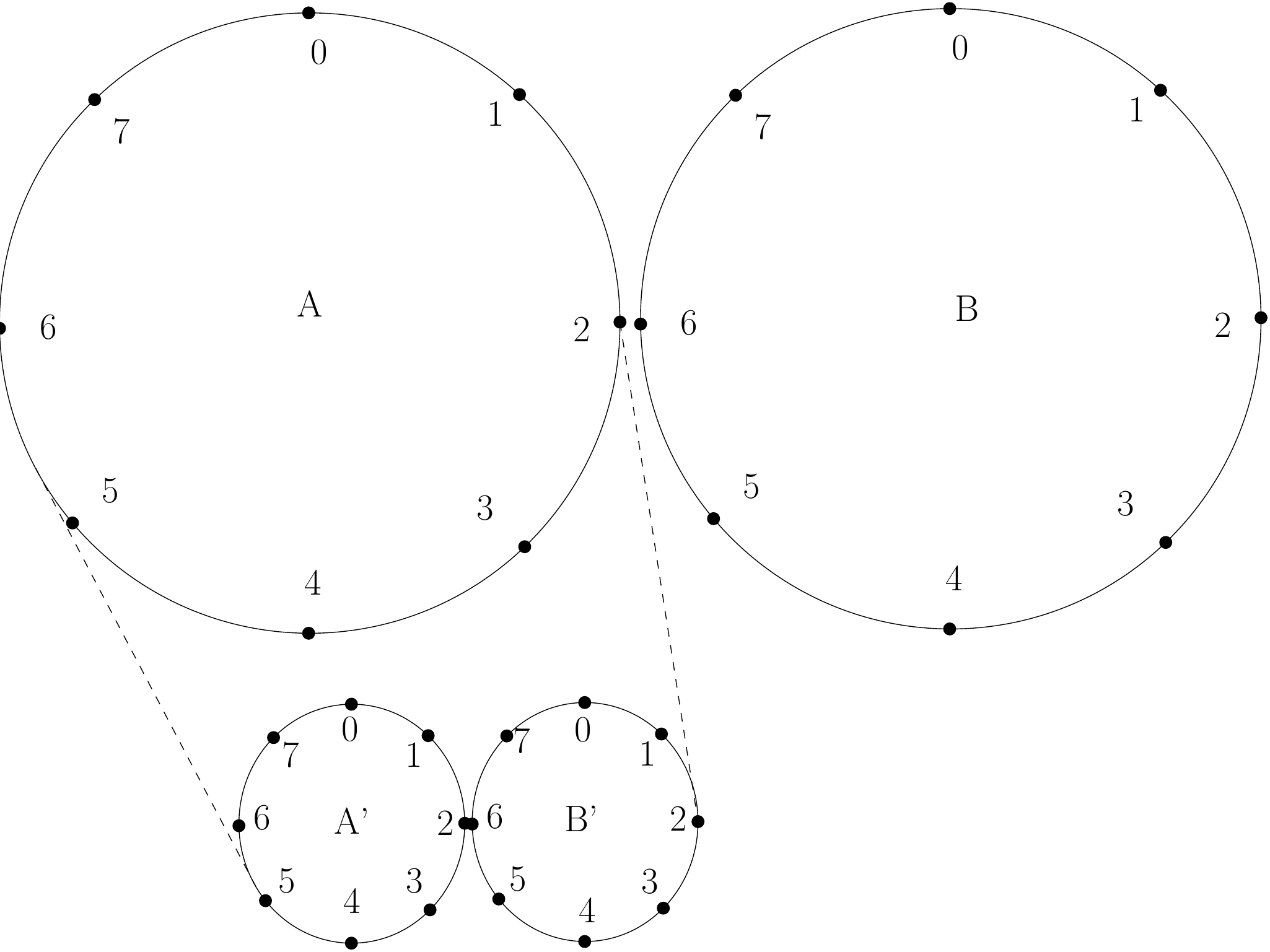}
 \caption{Conceptualization of the Recursive Partition Gray Code.
$A$ moves clockwise while $A \ne B$, at which point $B$
moves counter-clockwise.}
 \label{fig:rpgc}
\end{figure}

Figure \ref{fig:rpgc} shows a conceptualization of the Recursive Partition
Gray Code.  Wheels $A$ and $B$ represent the states in part $A$ and $B$
of the code, respectively.  When $A$ is incremented, it moves to the next
clockwise location, and when $B$ is decremented it moves to the next
counter-clockwise location.  And inside each wheel $A$ and $B$ is another set of
wheels.

The increment and decrement operations must partition
the bit strings identically; we assume w.l.o.g. that the $A$ partition contains
the first $n/2$ bits, and $B$ contains the remaining $n/2$ bits.
Pseudocode for the increment (\emph{RecurIncrementPow2}) and
decrement (\emph{RecurDecrementPow2}) operations follows.

\begin{algorithm}[H]
	\caption{RecurIncrementPow2}
	\label{proc:recur-inc2}
	\KwIn{$b[]$, an array of $n$ bits}
	\eIf{$n=1$}{
		\lIf{$b[0]=1$}{$b[0] \xleftarrow{} 0$}\;
		\lElse{$b[0] \xleftarrow{} 1$}\;
	}{
		Let $A = b[0...n/2-1]$\;
		Let $B = b[n/2...n-1]$\;
		\lIf{$A = B$}{$RecurDecrementPow2(B)$}\;
		\lElse{$RecurIncrementPow2(A)$}\;
	}
\end{algorithm}

\begin{algorithm}[H]
	\caption{RecurDecrementPow2}
	\label{proc:recur-dec2}
	\KwIn{$b[]$, an array of $n$ bits}
	\eIf{$n=1$}{
		\lIf{$b[0]=1$}{$b[0] \xleftarrow{} 0$}\;
		\lElse{$b[0] \xleftarrow{} 1$}\;
	}{
		Let $A = b[0...n/2-1]$\;
		Let $B = b[n/2...n-1]$\;
		\lIf{$A = B + 1$}{$RecurIncrementPow2(B)$}\;
		\lElse{$RecurDecrementPow2(A)$}\;
	}
\end{algorithm}

\begin{lemma}\label{lem:partcodedat2}
The Recursive Partition Gray Code algorithm
can be performed by a DAT to generate a Gray code of dimension $d$ when $d$
is a power of two.
\end{lemma}
\begin{proof}
We will recursively define a DAT $I_d$ that performs the
\emph{RecurIncrementPow2} operation, and a DAT $D_d$ that performs the
\emph{RecurDecrementPow2} operation.

Let $C_d$ be a DAT that compares two bit strings of dimension $d$ to determine
if they differ in at least one bit.
Each leaf of $C_d$ represents a single bit $i$ of the bit strings, such that $i$
is the first bit seen which differs.  Assume the DAT scans the bit strings for
differences from bit $0$ to bit $d-1$ in order.
Then number the leaves such that
leaf $C_{d,i}$ is reached when bit $i$ is the first bit seen that
differs between
the two bit strings.  Leaf $C_{d,d}$ is reached when there is no difference
between the two bit strings.

Let $M_d$ be a DAT that compares two bit strings of dimension $d$, to determine
if they differ in at least two bits.
As the root of $M_d$, use a $C_d$ DAT.  At leaves $C_{d,i}$, $0 \le i \le d-2$,
place another DAT $C_{d-i-1}$, that compares the last $d-i-1$ bits between the
two bit strings.  Call the leaves of these trees $C_{d,i,j}$ when they are the
$j$\emph{-th} leaf of the subtree rooted at
the $i$\emph{-th} leaf of the top-level $C_d$ DAT.
Then some leaf $C_{d,i,j}$, for $0 \le i \le d-2$ and $j \le d-i-1$, will be
reached when the two bit strings differ in at least two bits, while some leaf
$C_{d,i,j}$, for $i \ge d-1$ or $j \ge d-i$, will be reached when the two bit
strings have less than two bits which differ.  Call the first set of leaves
the 2-differ leaves, and the latter the 0-differ leaves.

$I_1$ and $D_1$ have a height of $1$, and are each constructed identically.
The root node reads the single bit and control transfers to a
child.  The leaf nodes flip the value of the single bit: if the bit was $1$ it
writes a $0$, and if the bit was $0$ it writes a $1$.

To construct a DAT for $I_d$, with $d > 1$, place a DAT $C_{d/2}$ as the root.
At each leaf except $C_{d/2,d/2}$, put the root node of an
$I_{d/2}$ DAT.  At the leaf $C_{d/2,d/2}$, place the root node of
a $D_{d/2}$ DAT.

To construct a DAT for $D_d$, with $d > 1$, place a DAT $M_{d/2}$ as the root,
which considers the entire bit string of dimension $d$.
At each 0-differ leaf in the $M_{d/2}$ DAT, place the root of a $I_{d/2}$ DAT,
which simulates an increment operation on the last $d/2$ bits,
with each leaf being the root of a $C_{d/2}$ DAT.  At each leaf of the
$C_{d/2}$ DATs, place the root of a $D_{d/2}$ DAT that operates also on the last
$d/2$ bits, with exception that the
highest rank leaf in each $C_{d/2}$ DAT is the root of a $I_{d/2}$ DAT
that operates on the first $d/2$ bits instead.
Finally, at each 2-differ leaf node, place a $D_{d/2}$ DAT that operates on
the last $d/2$ bits.

At the end of this process, look at each path from the root to a leaf in the
final DAT.  Let $i$ and $j$ be nodes on any such path such that $i$ and $j$
both read the same bit of the input string.  Assume without loss of generality
that $j$ is in the left subtree of $i$.  Then remove $j$ and its right subtree
from the DAT, replacing $j$ with its left child.  Continue this process until
there are no such $i$ and $j$ in the DAT.
\end{proof}

\begin{lemma}\label{lem:partcode2len}
For a dimension $d \ge 1$, the \texttt{RecurIncrementPow2} algorithm
generates a Gray code of dimension $d$ with length $2^d$.
\end{lemma}
\begin{proof}
The proof is by induction on $d$.

Let $d=1$, then the \texttt{RecurIncrementPow2} algorithm flips the bit twice,
creating a Gray code of length $2 = 2^1 = 2^d$.  The same is true for the
\texttt{RecurDecrementPow2} algorithm.

Then assume it is true for dimensions less than $d$.
We will show that for dimension $d$, the
\texttt{RecurIncrementPow2} algorithm generates a Gray code of length $2^d$.

Let $|X|$ be the dimension of a bit string $X$.

If $d$ is even, and $A=B$ initially, then the algorithm starts by
decrementing $B$.  For each decrement of $B$, the algorithm
will increment $A$ $2^{d/2}-1$ times in order to make them equal again.
This comes from the fact that $|A|=d/2 < d$ and
thus the bits move through $2^{d/2}$
unique states.

$B$ is decremented $2^{d/2}$ times before it reaches its initial state again,
since $|B|=d/2<d$.  Since only one of $A$ and $B$ is changed at each step,
the total number of bit strings generated is
equal to the number of times $B$ is decremented plus the number of times $A$
is incremented.  This is $2^{|B|} + 2^{|B|} \cdot (2^{|A|}-1) =
2^{d/2} + 2^{d/2}(2^{d/2}-1) = 2^{d/2} + 2^d-2^{d/2} = 2^d$.
The same holds for \texttt{RecurDecrementPow2} by a symmetric argument.

At each step if the algorithm, bits are read but not written, except in the
base case, where $d=1$.
Each generating step recurses on one sub problem, thus only one bit is
written, and the resulting code is a Gray code.

Therefore, the \texttt{RecurIncrementPow2} algorithm generates a Gray code of
dimension $d$ with length $L=2^d$.
\end{proof}

\begin{thm}\label{thm:partcode2}
Let $d \ge 2$ be a power of two.
There exists a DAT that generates a Gray code of
dimension $d$ and length $L=2^d$, where
generating the next bit string requires reading on average
no more than $4\log d$ bits
of the current string.  In the worst case, $d$ bits are read, and
only $1$ bit is written.
\end{thm}

\begin{proof}\label{pf:partcode2}
By Lemma \ref{lem:partcodedat2}, we can construct a DAT that performs the
\emph{RecurIncrementPow2} and \emph{RecurDecrementPow2} operations and generates
a Recursive Partition Gray code of dimension $d$.

Since the RPGC has length $L=2^d$,
the algorithm will be executed once for each possible bit string
of dimension $d$.  Based on this observation,
we bound the average number of bits read
by studying the expected number of bits read given a random bit string of
dimension $d$.
The proof is by induction on $d$.
For the base case $d=2$, in the worst case we read at most $2$ bits,
so the average number of bits read in a random bit string
is at most $2 \le 4\log d$. Then we assume our claim is
true for a random bit string of dimension $d/2$.

We define $|X|$ to denote the dimension of a bit string $X$.
Let $C(A,B)$ be the number of bits read to determine whether or not $A=B$,
where $A$ and $B$ are bit strings and $|A| = |B|$.
Let $I(X)$ be the number of bits read to increment the bit string $X$.
Let $D(X)$ be the number of bits read to decrement the bit string $X$.
Note that since we are working in the DAT model, we read any bit
at most one time, and $D(X) \le |X|$.

To finish the proof, we need to show that $\mathrm{E}[I(X)] \le 4\log d$, when
$X$ is a random bit string of dimension $d$.

We can determine the expected value of $C(A,B)$ as follows.
$C(A,B)$ must
read two bits at a time, one from each of $A$ and $B$,
and compare them, only until
it finds a pair that differs.  Given two random bit strings, the probability
that bit $i$ is the first bit that differs between the two strings is
$1/2^{i+1}$.  If the two strings differ in bit $i$, then the function will
read exactly $i+1$ bits in each string.
If $|A| = |B| = n/2$, then the expected value of $C(A,B)$ is
\[
\mathrm{E}[C(A,B)]
 = 2\sum_{i=0}^{n/2-1}\frac{i+1}{2^{i+1}}
 = 2\sum_{i=1}^{n/2}\frac{i}{2^i} = 2\left(\frac{2^{n/2+1} - n/2 - 2}{2^{n/2}}\right)
 = 2\left(2 - \frac{n/2+2}{2^{n/2}}\right)
\].

Let $X = AB$, and $|A| = |B| = d/2$.  Then $|X| = d$.

For a predicate $P$, we define $\mathbf{1}_P$ to be the indicator random variable
whose value is $1$ when $P$ is true, and $0$ otherwise.

Note that $I(A)$ is independent of $\mathbf{1}_{A=B}$ and $\mathbf{1}_{A \ne B}$.
This is because the relation between $A$ and $B$ has no effect
on the distribution of $A$ (which remains uniformly distributed among all
bit strings of dimension $d/2$).  The same is true of $I(B)$, $D(A)$ and $D(B)$.

The \emph{RecurIncrementPow2} operation only performs one increment or decrement
action, depending on the condition $A = B$, thus the expected number of bits
read by $I(X)$ is
\begin{align*}
\mathrm{E}[I(X)] & = \mathrm{E}[C(A,B)] + \mathrm{E}[\mathbf{1}_{A=B}D(B)] + \mathrm{E}[\mathbf{1}_{A \ne B}I(A)] \\
 & \le 2\left(2 - \frac{d/2 + 2}{2^{d/2}}\right) + \frac{1}{2^{d/2}}\frac{d}{2} + (1-\frac{1}{2^{d/2}})\mathrm{E}[I(A)] \\
& \le 4 - \frac{d/2+4}{2^{d/2}} + 4\log(d/2) \\
& \le 4 + 4\log d  - 4\log 2 \\
& = 4\log d \text{ ,}
\end{align*}
as required.

Last, in the worst case when $A=B$,
the comparison test will require reading all bits
in $A$ and $B$, and thus all $d$ bits in the code.
\end{proof}

Now consider the case when $d$ is not a power of two.
When the increment operation is given a bit string with an even number of bits,
the algorithm is the same as in the power of two case.  However, when it is
given a bit string with an odd number of bits,
it uses the first bit as a direction bit.  While the direction bit is $0$, the
increment operation will recursively increment the remaining bits,
until they reach their state
of highest rank.  At that point, the increment operation flips the
direction bit to $1$.  From then on, while the direction bit is $1$,
the increment operation will recursively decrement the remaining bits,
until they reach their state of minimum rank.
Then the increment operation would flip the direction bit back to $0$,
reaching its initial state.  This generates all possible states for the bit
string.

The decrement operation functions similarly to the increment operation
when given a bit string with an odd number of bits, except that it
decrements the remaining bits when the direction bit is $0$ and increments
them when the direction bit is $1$.

Pseudocode for the increment (\texttt{RecurIncrement}) and
decrement (\emph{RecurDecrement}) operations for the more
general scenario follows.

\begin{algorithm}[H]
	\caption{RecurIncrement}
	\label{proc:recur-inc}
	\KwIn{$b[]$, an array of $n$ bits}
	\eIf{$n=1$}{
		\lIf{$b[0]=1$}{$b[0] \xleftarrow{} 0$}\;
		\lElse{$b[0] \xleftarrow{} 1$}\;
	}{
		\eIf{$n \text{ \emph{mod} } 2 = 1$}{
			Let $W = b[1...n-1]$\;
			\eIf{$b[0] = 0$}{
				\tcp{Is $W$ in its maximal ranked state $1000...0$?}
				\lIf{$rank(W) = 2^{n-1}-1$}{$b[0] \xleftarrow{} 1$}\;
				\lElse{$RecurIncrement(W)$}\;
			}{
				\tcp{Is $W$ in its minimal ranked state $000...0$?}
				\lIf{$rank(W) = 0$}{$b[0] \xleftarrow{} 0$}\;
				\lElse{$RecurDecrement(W)$}\;
			}
		}{
			Let $A = b[0...n/2-1]$\;
			Let $B = b[n/2...n-1]$\;
			\lIf{$A = B$}{$RecurDecrement(B)$}\;
			\lElse{$RecurIncrement(A)$}\;
		}
	}
\end{algorithm}

\begin{algorithm}[H]
	\caption{CompareInc}
	\label{proc:comp-inc}
	\KwIn{$A[]$, an array of $n$ bits; $B[]$, an array of $n$ bits}
	\tcp{Returns \emph{true} if incrementing $B$ would put the bits of $B$ in
	     the same state as the bits in $A$}
	\eIf{$n=1$}{
		return $A[0] \ne B[0]$\;
	}{
		Let $A_1 = A[0...n/2-1]$\;
		Let $A_2 = A[n/2...n-1]$\;
		Let $B_1 = B[0...n/2-1]$\;
		Let $B_2 = B[n/2...n-1]$\;
		\eIf{$B_1 = B_2$}{
			\lIf{$A_1 \ne B_1$}{ return \emph{false} }\;
			\lElse{return $CompareInc(B_2, A_2)$}\;
		}{
			\lIf{$A_2 \ne B_2$}{ return \emph{false} }\;
			\lElse{return $CompareInc(A_1, B_1)$}\;
		}
	}
\end{algorithm}

\begin{algorithm}[H]
	\caption{RecurDecrement}
	\label{proc:recur-dec}
	\KwIn{$b[]$, an array of $n$ bits}
	\eIf{$n=1$}{
		\lIf{$b[0]=1$}{$b[0] \xleftarrow{} 0$}\;
		\lElse{$b[0] \xleftarrow{} 1$}\;
	}{
		\eIf{$n \text{ \emph{mod} } 2 = 1$}{
			Let $W = b[1...n-1]$\;
			\eIf{$b[0] = 0$}{
				\tcp{Is $W$ in its minimal ranked state $000...0$?}
				\lIf{$rank(W) = 0$}{$b[0] \xleftarrow{} 1$}\;
				\lElse{$RecurDecrement(W)$}\;
			}{
				\tcp{Is $W$ in its maximal ranked state $1000...0$?}
				\lIf{$rank(W) = 2^{n-1}-1$}{$b[0] \xleftarrow{} 0$}\;
				\lElse{$RecurIncrement(W)$}\;
			}
		}{
			Let $A = b[0...n/2-1]$\;
			Let $B = b[n/2...n-1]$\;
			\lIf{$CompareInc(A, B)$}{$RecurIncrement(B)$}\;
			\lElse{$RecurDecrement(A)$}\;
		}
	}
\end{algorithm}

\begin{lemma}\label{lem:partcodelen}
For a dimension $d \ge 1$, the Recursive Partition Gray Code algorithm
generates a Gray code of dimension $d$ with length $2^d$.
\end{lemma}
\begin{proof}
The proof is by induction on $d$.

Let $d=1$, then the \texttt{RecurIncrement} algorithm flips the bit twice,
creating a Gray code of length $2 = 2^1 = 2^d$.  The same is true for the
\texttt{RecurDecrement} algorithm.

Then assume it is true for dimensions less than $d$.
We will show that for dimension $d$, the
\texttt{RecurIncrement} algorithm generates a Gray code of length $2^d$.

If $d$ is even, and $A=B$ initially, then the algorithm starts by
decrementing $B$.  For each decrement of $B$, the algorithm
will increment $A$ $2^{d/2}-1$ times in order to make them equal again.
This comes from the fact that $|A|=d/2 < d$ and
thus the bits move through $2^{d/2}$
unique states.

$B$ is decremented $2^{d/2}$ times before it reaches its initial state again,
since $|B|=d/2<d$.  Since only one of $A$ and $B$ is changed at each step,
the total number of bit strings generated is
equal to the number of times $B$ is decremented plus the number of times $A$
is incremented.  This is $2^{|B|} + 2^{|B|} \cdot (2^{|A|}-1) =
2^{d/2} + 2^{d/2}(2^{d/2}-1) = 2^{d/2} + 2^d-2^{d/2} = 2^d$.
The same holds for \texttt{RecurDecrement} by a symmetric argument.

If $d$ is odd, then the algorithm will increment the last $d-1$ bits
until they reach their maximally ranked state.  Since $d-1 < d$, this will
generate $2^{d-1} - 1$ bit strings.
Next, the algorithm will flip the first bit, one bit string, and
decrement the last $d-1$ bits until they reach their initial state.  This
generates another $2^{d-1}-1$ bit strings. Finally, it flips the first bit
back to its initial state as well.  In total, this generates
$2^{d-1}-1 + 1 + 2^{d-1}-1 + 1 = 2^d$ bit strings.
The \texttt{RecurDecrement} algorithm performs the same operations in
a different order, producing the same number of bit strings.

At each step if the algorithm, bits are read but not written, except in the
base case, where $d=1$.
Each generating step recurses on one sub problem, thus only one bit is
written, and the resulting code is a Gray code.

Therefore, the Recursive Partition Gray Code algorithm generates a Gray code of
dimension $d$ with length $L=2^d$.
\end{proof}

\begin{thm}\label{thm:partcode}
Let $d \ge 2$.
There exists a DAT that generates a Gray code of
dimension $d$ and length $L=2^d$, where
generating the next bit string requires reading on average
no more than $6\log d$ bits
of the current string.  Only $1$ bit is written in the worst case,
and $d$ bits are read.
\end{thm}

\begin{proof}\label{pf:partcode}
From Lemma \ref{lem:partcodelen}, we know the length of a RPGC of dimension
$d$ is exactly $2^d$.
Then, as in the proof of Theorem \ref{thm:partcode2},
the algorithm will be executed once for each possible bit string
of dimension $d$.  As such, we bound the average number of bits read
by studying the expected number of bits read given a random bit string of
dimension $d$.
The proof is by induction on $d$.  We will show that
the expected number of bits read to increment is at most $5.623\log d$.
In some cases, incrementing requires decrementing a substring of the code,
and we will show that in these cases, the expected number of
bits read by a decrementing step is at most $7.746 \log d$ bits.

For the base case $d=2$, in the worst case we read at most $2$ bits
to perform \texttt{RecurIncrement},
so the average bits read is at most $2 \le 5.623 \log d$.
Similarly to perform \emph{RecurDecrement}, we read at most $2$ bits,
and thus read on average at most $2 \le 7.746\log d$.

Then we assume that these both hold
for a random bit string of dimension less than $d$.
Let $I(X)$ and $D(X)$ be the number of bits read to
increment or decrement the bit string $X$, respectively.
To finish the proof, we will show that $\mathrm{E}[I(X)] \le 5.623\log d$,
when $X$ is a random bit string of dimension $d$.

We define $|X|$ to denote the dimension of the bit string $X$,
and $X^{-t}$ refers to a substring of $X$ with dimension $|X|-t$.
Let $C(A,B)$ be the number of bits read to determine whether or not $A=B$,
where $A$ and $B$ are bit strings and $|A| = |B| = n/2$.
Let $M(A,B)$ be the number of bits read by the \emph{CompareInc} algorithm,
which determines whether or not $A=B+1$,
where $A$ and $B$ are bit strings and $|A| = |B| = n/2$.
Note that since we are working in the DAT model, we read any bit
at most one time.
And finally, let $I=3.5$ and $D=7.746$.

The expected value of $C(A,B)$ is given in the proof of
Theorem \ref{thm:partcode2} as 
$\mathrm{E}[C(A,B)] = 2\left(2 - \frac{2+n/2}{2^{n/2}}\right)$,
and we can determine the expected value of $M(A,B)$ as follows.
$M(A,B)$ compares two bit strings against
a third, all of equal dimension, $n/4$.  There are four possible outcomes
for the \emph{CompareInc} function.  We examine the cost and probability of
each to determine the expected value of $M(A,B)$.

When $B_1 = B_2 \ne A_1$, \emph{CompareInc} returns \emph{false}.  In this case
$B_1$ is compared against $B_2$ and $A_1$.  Since $B_1 = B_2$, we read all
$n/2$ bits in $B_1$ and $B_2$.  Since $A_1$ differs from the others,
we expect to read
$\sum_{i=1}^{n/4}\frac{i}{2^i} = 2 - \frac{2+n/4}{2^{n/4}}$ bits of $A_1$,
by the same argument as made for $C(A,B)$.
Thus we read at most $2+n/2$ bits on average in
this case, which happens with probability
$
 \frac{(2^{n/4})(2^{n/4})(2^{n/4}-1)}{2^n} = \frac{2^{3n/4}-2^{n/2}}{2^n}
 = \frac{2^{n/4} - 1}{2^{n/2}}
$.

When $B_1 \ne B_2$ and $B_2 \ne A_2$, \emph{CompareInc} returns \emph{false}
again.  In this case $B_2$ is compared against $B_1$ and against $A_2$.  If the
two comparisons read the bits of $B_2$ in the same order, the second comparison
will read many of the same bits as the first.  In the DAT model, this means we
can count far fewer bits read on average for the second comparison.
The expected number of bits read in each of $B_1$ and $A_2$ is
$2 - \frac{2+n/4}{2^{n/4}}$.
The number of bits read in $B_2$ is the max of the number of bits
read in $B_1$ and $A_2$.  Consider an equivalent problem.  Given two random
bit strings of dimension $k$, if we examine bit positions in both strings
together, count the number of positions we expect to look at
to find at least one $1$ in each string.
The probability that position $i$, $1 \le i \le k$, is the
first position where we have seen a $1$ at some
position $j \le i$ in each string is
$2\frac{2^{2(k-i)}(2^i-1)}{2^{2k}} - \frac{2^{2(k-i)}}{2^{2k}} =
2\frac{2^i-1}{2^{2i}} - \frac{1}{2^{2i}}$.
The first term counts the number of strings with the first $i-1$ bits set to $0$
and the $i$-th bit set to $1$ for one string,
and at least one $1$ somewhere in the first $i$ bits of the other.
It is multiplied by two, since this can occur with a $1$ in position $i$ for
either of the two strings.
The second term keeps us from double counting the case where the first $1$
in both strings occurs at position $i$.  The expected number of bit positions
read is then
\[
 \sum_{i=1}^{k}i(2\frac{2^i-1}{2^{2i}} - \frac{1}{2^{2i}}) =
 2\sum_{i=1}^{k}\left( \frac{i}{2^i} - \frac{3}{2}\frac{i}{2^{2i}} \right)
 \le 8/3
 \text{ ,}
\]
which is the expected number of bits read in $B_2$.
The above summation converges to $8/3$ in the limit $i \to \infty$,
and so we can upper bound it by this value.
Thus, in this case,
the average number of bits we read is at most $20/3$.  This scenario occurs
with probability
$\frac{(2^{n/4})^2(2^{n/4}-1)^2}{2^n} = \frac{2^{n/2}-2^{1+n/4}+1}{2^{n/2}}
= 1 - \frac{2^{1+n/4}-1}{2^{n/2}}$.

In the other two cases, \emph{CompareInc} recurses on a problem of size $n/2$.
Since there are $n$ bits in the input to the $M(A,B)$, we can bound the average
number of bits read in this case by $n$, which happens with probability
$\frac{(2^{n/4})(2^{n/4})}{2^n} + \frac{(2^{n/4})^2(2^{n/4}-1)}{2^n}
 = \frac{1}{2^{n/4}}$.

By combining these four cases, the expected value of $M(A,B)$ is
\[
 \mathrm{E}[M(A,B)]
 = \left(\frac{2^{n/4} - 1}{2^{n/2}}\right)(2+n/2) +
   \left(1 - \frac{2^{1+n/4}-1}{2^{n/2}}\right)(20/3) +
   \left(\frac{1}{2^{n/4}}\right)n
 \le 7.507
 \text{ ,}
\]
for $n \ge 2$.
This bound comes from maximizing the function, where
according to Maple, the function has a maximum value $\approx 7.506883004$,
for $n \ge 2$.

For a predicate $P$, we define $\mathbf{1}_P$ to be the indicator random
variable whose value is $1$ when $P$ is true, and $0$ otherwise.

Let $A$ and $B$ be two random bit strings of equal dimension.
Note that $I(A)$, $I(B)$, $D(A)$, and $D(B)$ are all independent of
$\mathbf{1}_{A=B}$ and $\mathbf{1}_{A = B+1}$.
This is because the relation between $A$ and $B$ has no effect
on the distribution of $A$ or $B$
(which each remain uniformly distributed among all bit strings of
their dimension).

Throughout the following cases, we assume that the first bit of $A$ or $B$,
if we recurse on $A$ or $B$ respectively, will always be read in the next
recursive step, so we are able to read the bit for free in the current step.
This is true, since we count the first bit explicitly in each inductive step,
and the base case reads all of its bits, including the first.

Let $X$ be an input string of dimension $d$.  We consider the cases
when $d$ is an even or odd number separately.  First, assume that $d$ is
an even number.

The \texttt{RecurIncrement} operation only performs one increment or decrement
action, depending on the condition $A = B$.  We bound the cost
of $D(B)$ here by $d/2$ which is an upper bound for the DAT model.
The comparison in $C(A,B)$ will read the first bit of $A$ and $B$, so at least
one bit will be double counted, which we can subtract.
Thus the expected number of bits read by $I(X)$, when $|X|$ is even, is
\begin{align*}
\mathrm{E}[I(X)] & = \mathrm{E}[C(A,B)] - 1 + \mathrm{E}[\mathbf{1}_{A=B}D(B)] + \mathrm{E}[\mathbf{1}_{A \ne B}I(A)] \\
& \le 2\left(2 - \frac{d/2 + 2}{2^{d/2}}\right) - 1 + \frac{1}{2^{d/2}}\frac{d}{2} + (1-\frac{1}{2^{d/2}})\mathrm{E}[I(A)] \\
& = 3 - \frac{d/2+4}{2^{d/2}} + \mathrm{E}[I(A)] \\
& \le 3 + I\log(d/2) \\
& \le I\log d
 \text{ ,}
\end{align*}
because $I = 3.5 \ge 3$.

We also examine the cost of the \emph{RecurDecrement} operation, as we will
need it for the case when $d$ is odd.
The \emph{RecurDecrement} operation performs one decrement or increment
action, depending on the condition $A = B + 1$.
We make sure that we count both the first bit of $A$ and $B$ so that we can
subtract one bit as being double counted.  The first bit of $B$ is always read
by $M(A,B)$, but the first bit of $A$ is not read when $B_1 \ne B_2$.
We bound the cost
of $I(B)$ here by $d/2$, giving us the expected number of
bits read by $D(X)$, when $|X|$ is even, as
\begin{align*}
\mathrm{E}[D(X)] & = \mathrm{E}[M(A,B)] + \mathrm{E}[\mathbf{1}_{B_1 \ne B_2}1] - 1 + \mathrm{E}[\mathbf{1}_{A=B+1}I(B)] + \mathrm{E}[\mathbf{1}_{A \ne B+1}D(A)] \\
& \le 7.507 + \frac{1}{2^{d/4}} - 1 + \frac{d/2}{2^{d/2}} + \mathrm{E}[D(A)] \\
& \le 6.507 + \frac{2^{d/4}+d/2}{2^{d/2}} + \mathrm{E}[D(A)] \\
& \le 7.715 + D\log(d/2) \\
& \le D\log d
\text{ ,}
\end{align*}
because $D = 7.746 \ge 7.715$. Elementary algebraic manipulation shows that 
$\frac{2^{d/4}+d/2}{2^{d/2}}$ is at most $\frac{\sqrt{2}+1}{2} \le 1.208$.

Now consider the case when $d$ is an odd number.
Let $X = xAB$, where $|x| = 1$ and $|A| = |B| = (d-1)/2$.
We assume that the bit $x$ can be read for free, which we will later
show is the case.

When $x=0$, the \texttt{RecurIncrement} operation checks to see that
$AB \ne 100...0$, and increments $AB$ when it is true, otherwise it stops and
reads no more bits.  Let $a_i$ and $b_j$ be
the $i$\emph{-th} and $j$\emph{-th} bits of $A$
and $B$, respectively.  We assume that the comparison of $AB$ to the bit string
$100...0$ is done by looking at the bits of $A$ and of $B$ in an
alternating sequence, such that the first bit read is $a_0$, then $b_0$,
followed by $a_1$ and $b_1$, and so on.

Consider the four cases for the possible values of $a_0$ and $b_0$, the first
bits read.

\textbf{Case $a_0=0$, $b_0=0$:} Then we know $AB \ne 100...0$ and $AB$ will
be incremented.  The first step of incrementing $AB$ is to test $A=B$.
This test will read the first bit of $A$ and $B$, and so at least one bit
is double-counted and can be subtracted here.
Thus, in this case, the expected number of bits read will be at most
$2 - 1 +  \mathrm{E}[C(A^{-1},B^{-1})]
 + \mathrm{E}[\mathbf{1}_{A^{-1} \ne B^{-1}}I(A)] + \mathrm{E}[\mathbf{1}_{A^{-1}=B^{-1}}D(B)]
= 1 + 4 - 2\frac{2+(d/2-1)}{2^{d/2-1}}
 + \mathrm{E}[\mathbf{1}_{A^{-1} \ne B^{-1}}I(A)] + \mathrm{E}[\mathbf{1}_{A^{-1}=B^{-1}}D(B)]
= 5 - \frac{3+d/2}{2^{d/2-1}}
 + \mathrm{E}[\mathbf{1}_{A^{-1} \ne B^{-1}}I(A)] + \mathrm{E}[\mathbf{1}_{A^{-1}=B^{-1}}D(B)]$.

\textbf{Case $a_0=1$, $b_0=1$:} This case has the same analysis and
expected value as the previous case.

\textbf{Case $a_0=1$, $b_0=0$:} Then we must continue checking
if $AB = 100...0$.
This will read an expected $2 - (2+(d-2))2^{-(d-2)}$ bits.  We already know
the result of the test $A=B$
from the bits $a_0$ and $b_0$, so no more bits need to be read
in order to proceed.  And we can subtract one bit as we double count the first
bit of $A$ or $B$ once again.
Thus, in this case, the expected number of bits read will be at most
$2 - 1  + 2 - \frac{2+(d-2)}{2^{d-2}}
+ \mathrm{E}[\mathbf{1}_{A^{-1} \ne B^{-1}}I(A)] + \mathrm{E}[\mathbf{1}_{A^{-1}=B^{-1}}D(B)]
= 3 - \frac{d}{2^{d-2}} 
+ \mathrm{E}[\mathbf{1}_{A^{-1} \ne B^{-1}}I(A)] + \mathrm{E}[\mathbf{1}_{A^{-1}=B^{-1}}D(B)]$.

\textbf{Case $a_0=0$, $b_0=1$:} Then we know $AB \ne 100...0$ and also that
$A \ne B$, and we do not read any more bits to continue to the next recursive
step.  We have also double counted the first bit of $A$ or $B$ and can subtract
it. Thus, in this case, the expected number of bits read will be at most
$2 - 1 + \mathrm{E}[\mathbf{1}_{A^{-1} \ne B^{-1}}I(A)] + \mathrm{E}[\mathbf{1}_{A^{-1}=B^{-1}}D(B)]
=1 + \mathrm{E}[\mathbf{1}_{A^{-1} \ne B^{-1}}I(A)] + \mathrm{E}[\mathbf{1}_{A^{-1}=B^{-1}}D(B)]$.

We bound the number of bits read for $D(B)$ by $d/2$, as this is an upper bound
for the DAT model.
Because each of these cases occurs once out of every four executions, the
total expected value of $I(X)$ when $|X|$ is odd and $x=0$ is
\begin{eqnarray*}
\mathrm{E}[I_{x=0}(X)]
 &\le& \frac{1}{2}\left(5-\frac{3+d/2}{2^{d/2-1}}\right)
   + \frac{1}{4}\left(3-\frac{d}{2^{d-2}}\right)
   + \frac{1}{4}\left(1\right) \\
 &&+\ \mathrm{E}[\mathbf{1}_{A \ne B}I(A)] + \mathrm{E}[\mathbf{1}_{A=B}D(B)] \\
 &\le& \frac{14}{4} - \frac{3+d/2}{2^{d/2}} - \frac{d}{2^d} + \frac{d/2}{2^{d/2}} + \mathrm{E}[I(A)] \\
 &=& \frac{14}{4} - \frac{3}{2^{d/2}} - \frac{d}{2^d} + \mathrm{E}[I(A)] \\
 &\le& \frac{14}{4} + I\log(d/2) \\
 &=& I\log d
\text{ ,}
\end{eqnarray*}
because $I = 3.5 = 14/4$.

Next we consider when $x=1$.  In this scenario,
the \texttt{RecurIncrement} operation checks to see that
$AB \ne 000...0$, and decrements $AB$ when it is true, otherwise it stops and
reads no more bits.  Let $a_i$ and $b_j$ be
the $i$\emph{-th} and $j$\emph{-th} bits of $A$
and $B$, respectively.
In this case, we do the comparison of $AB$ to the bit string $000...0$ in a
different order.  We first look at the bit $b_0$, then $b_{n/2}$, followed
by $b_1$ and $b_{n/2+1}$.  We continue reading $B$ in this alternating order
until all bits have been
read.  Then we check the bits $a_{n/2}$, $a_{n/2+1}$, up to $a_{n-1}$, and
finally $a_0$, $a_1$, up to $a_{n/2-1}$.

Consider the four cases for the possible values of $b_0$ and $b_{n/2}$,
the first bits read.  And note that $M(A,B)$ always reads $b_0$ and $b_{n/2}$
so we do not count them outside of that function.

\textbf{Case $b_0=1$, $b_{n/2}=0$:} Then we know that $AB \ne 000...0$.  The
decrement operation requires checking that $A \ne B+1$, using the
\emph{CompareInc} function.  This number of bits read by this function is
$M(A,B)$.  We also count the first bit of $A$
if $B_1 \ne B_2$ and it is not read by $M(A,B)$, so
that it is double-counted and we can subtract it.
Thus, in this case, the expected number of bits read will be at most
$\mathrm{E}[M(A,B)] + \mathrm{E}[\mathbf{1}_{B_1 \ne B_2}1] - 1
 + \mathrm{E}[\mathbf{1}_{A \ne B+1}D(A)] + \mathrm{E}[\mathbf{1}_{A=B+1}I(B)]
\le 7.507 + \frac{1}{2^{d/4}} - 1
 + \mathrm{E}[\mathbf{1}_{A \ne B+1}D(A)] + \mathrm{E}[\mathbf{1}_{A=B+1}I(B)]
\le 7.215
 + \mathrm{E}[\mathbf{1}_{A \ne B+1}D(A)] + \mathrm{E}[\mathbf{1}_{A=B+1}I(B)]$.
This follows from the fact that $\frac{1}{2^{d/4}}$ is a decreasing function
for $d \ge 2$, and its maximum value,
at $d=2$, is $2^{-1/2} \le 0.708$.

\textbf{Case $b_0=0$, $b_{n/2}=1$:} This case has the same analysis and expected
value as the previous case.

\textbf{Case $b_0=1$, $b_{n/2}=1$:} This case also has the same analysis and
expected value as the previous two cases.

\textbf{Case $b_0=0$, $b_{n/2}=0$:} Then we need to continue checking if
$AB \ne 000...0$, and if it is true, $AB$ will
be decremented.
We note that this check reads bits in the same order as $M(A,B)$ compares
the bit strings $A$ and $B$.  While the check continues to find all $0$ bits,
the function \emph{CompareInc} would also find equality, so we know that
it would read at least as many bits as we check to find a bit set to $1$.
This holds until $3n/4$ bits have been read, which is the most that $M(A,B)$
will read.  When these bits are all $0$, we must continue reading the last
$n/4$ bits to compare them against $000...0$.  This happens for
$2^{n/4}$ of the possible $2^n$ inputs, thus with probability $2^{-3n/4}$.
To compare the last $n/4$ bits against $000...0$, we expect to read
$2-\frac{2+n/4}{2^{n/4}}$ bits.
We also make sure the first bit of $A$ is double-counted, as when $B_1 \ne B_2$,
it will not be read by $M(A,B)$ or by the comparison to $000...0$.
Thus, in this case, the expected number of bits read will be
at most
$\mathrm{E}[M(A,B)] + \mathrm{E}[\mathbf{1}_{B_1 \ne B_2}1] - 1 + \frac{1}{2^{3n/4}}\left(2-\frac{2+d/4}{2^{d/4}}\right)
 + \mathrm{E}[\mathbf{1}_{A \ne B+1}D(A)] + \mathrm{E}[\mathbf{1}_{A=B+1}I(B)]
\le 7.507 + \frac{1}{2^{d/4}} - 1 + \frac{2}{2^{3d/4}} - \frac{2+d/4}{2^d}
 + \mathrm{E}[\mathbf{1}_{A \ne B+1}D(A)] + \mathrm{E}[\mathbf{1}_{A=B+1}I(B)]$.

We bound the number of bits read for $I(B)$ by $d/2$, as this is an upper bound
for the DAT model.
If $|A|$ is even, we can bound the expected cost of $D(A)$ as shown above.
When $|A|$ is odd, then $D(A)$ will have the same expected number of bits read
as $I(A)$, since they both increment half the time, and decrement half the time.
Thus, in both cases,
we are able to bound $\mathrm{E}[D(A)]$ by $7.746 \log (d/2)$ by induction.
Because each of these cases occurs once out of every four executions, the
total expected value of $I(X)$ when $|X|$ is odd and $x=1$ is
\begin{eqnarray*}
\mathrm{E}[I_{x=1}(X)]
&\le& \frac{3}{4}\left(7.215\right)
    + \frac{1}{4}\left(7.507-1+\frac{1}{2^{d/4}} + \frac{2}{2^{3d/4}} - \frac{2+d/4}{2^d}\right) \\
    &&+\ \mathrm{E}[\mathbf{1}_{A \ne B+1}D(A)] + \mathrm{E}[\mathbf{1}_{A=B+1}I(B)] \\
&\le& 7.038 + \frac{1}{4}\left(\frac{1}{2^{d/4}} + \frac{2}{2^{3d/4}} - \frac{2+d/4}{2^d}\right)
    + \frac{d/2}{2^{d/2}} + \mathrm{E}[\mathbf{1}_{A \ne B+1}D(A)] \\
&=& 7.038 + \frac{1}{4}\left(\frac{1}{2^{d/4}} + \frac{2}{2^{3d/4}} - \frac{2+d/4}{2^d} + \frac{2d}{2^{d/2}}\right)
    + \mathrm{E}[\mathbf{1}_{A \ne B+1}D(A)] \\
&\le& 7.038 + \frac{2.832}{4} + \mathrm{E}[D(A)] \\
&\le& 7.746 + D\log(d/2) \\
&=& D\log d
 \text{ ,}
\end{eqnarray*}
because $D = 7.746$.  This follows from maximizing the function
$\frac{1}{2^{d/4}} + \frac{2}{2^{3d/4}} - \frac{2+d/4}{2^d} + \frac{4\cdot d/2}{2^{d/2}}$,
which according to Maple,
has a maximum value $\approx 2.831931894$ for $d \ge 2$.

Last, we must consider the bit $x$.  While generating the entire Gray code,
the bit $x$ will be $0$ for half of the time, and $1$ the other half.
Thus $I_{x=0}(X)$ will execute half of the time when $|X|$ is odd, and
$I_{x=1}(X)$ will execute the other half.
Therefore, the expected value of $I(X)$ when $|X|$ is odd is
\begin{align*}
\mathrm{E}[I(X)]
 &\le \mathrm{E}[\mathbf{1}_{x=0}I_{x=0}(X)] + \mathrm{E}[\mathbf{1}_{x=1}I_{x=1}(X)] \\
 &\le \frac{1}{2}(I\log d) + \frac{1}{2}(D\log d) \\
 &= \frac{1}{2}(3.5 \log d) + \frac{1}{2}(7.746 \log d) \\
 &\le 5.623\log d
 \text{ ,}
\end{align*}
because $I = 3.5$ and $D = 7.746$.  From our initial choice of $I$ and $D$,
all of the above holds, and we have shown that in all cases, for a bit string
$X$ of dimension $d$, $\mathrm{E}[I(X)] \le 5.623\log d$.

This satisfies the inductive proof, however, we did not count the cost of
reading the bit $x$ when the input string is odd.  Note that
when we recurse on a substring $A$ or $B$ of $X$, we have compared $A=B$
(to increment) or $A=B+1$ (to decrement), each of which requires reading at
least one bit of $A$ and $B$.  Since this bit will be $x$ when $A$ or $B$ is
odd, we will have always already read the bit $x$ during the
previous recursive step.  Therefore, the bit $x$ can be read for free whenever
we are inside a recursive call.  Only at the first level of recursion must
we count reading the bit $x$.
For any $d$, the algorithm's recursion eventually reaches the base case of
$d=2$.  In this case, the worst case number of bits read will be $2$, but
we count it as $6\log d = 6$.  This leaves $6-2=4$ bits counted in our average
that we did not actually read.  These bits more than account for the one bit $x$
at the top level of the recursion.

Therefore, to generate the next bit string
in the Recursive Partition Gray Code, given an input string $X$,
the \texttt{RecurIncrement} algorithm will read on average
$\mathrm{E}[I(X)] \le 6\log d$ bits.
\end{proof}

\section{Composite code construction}
\label{subsec:composite}

In this section, we show a method to take a code with space
efficiency $1$ that reads on
average $r$ bits, and construct a new, larger, code that reads
on average $O(\log\log r)$ bits, while maintaining the same space efficiency,
but increasing the worst-case number of bits written by one.

\begin{lemma}\label{lem:composite}
Let $d \ge 1$, $r \ge 3$, $p \ge 1$, $w \ge 1$ be integers.  Assume we have a
DAT for a code of dimension $d$, that generates $L=2^d$ bit strings,
such that the following holds: Given a bit string of length $d$, generating
the next bit string in the code
requires reading no more than $r$ bits on average, reads $p$ bits in the
worst case, and writes at most $w$ bits in the worst case.

Then there is a space-optimal DAT for a code of dimension
$d + \lceil \log r \rceil$, where
generating each bit string from the previous one requires
reading no more than $6\log \lceil\log r\rceil + 3$ bits on average,
reading $p + \lceil \log r \rceil$ bits in the worst case,
and writing at most $w + 1$ bits.
That is, the average number of bits read decreases
from $r$ to $O(\log\log r)$, while the worst-case number of bits written
increases by one.
\end{lemma}

\begin{proof}\label{pf:composite}
We are given a DAT $A$ that generates a code of dimension $d$.
The DAT, $A$, requires reading on
average no more than $r$ bits, and writing at most $w$ bits to generate
each bit string.
 We construct a DAT $B$ for the
Recursive Partition Gray Code of dimension $d'$,
such as described in Section \ref{subsec:partition}.
The DAT, $B$, requires reading $\le 6\log d'$ bits on average to generate the
next state, and requires writing only $1$ bit in the worst case.

We construct a new DAT from the two DATs of dimension $d$ and $d'$.
The combined DAT generates bit strings of dimension $d + d'$.  The last $d'$
bits of the combined code, when updated, will cycle through the
code generated by $B$.  The first $d$ bits, when updated, will
cycle through the code generated by $A$.

The DAT
initially moves the last $d'$ bits through $2^{d'}$ states according to the
rules of $B$.  When it leaves this final state, to generate the initial bit
string of $B$ again, the DAT also moves the first $d$ bits to their next state
according to the rules of $A$.

Figure \ref{fig:composite} shows a conceptualization of the Composite
code.  Wheels $A$ and $B$ represent the states in part $A$ and $B$
of the code, respectively.  $B$ moves clockwise around its wheel with each
increment operation.  When $B$ moves from its highest state back to $0$,
then $A$ also moves one step in the clockwise direction.

\begin{figure}
 \centering
 \includegraphics[width=0.75\linewidth]{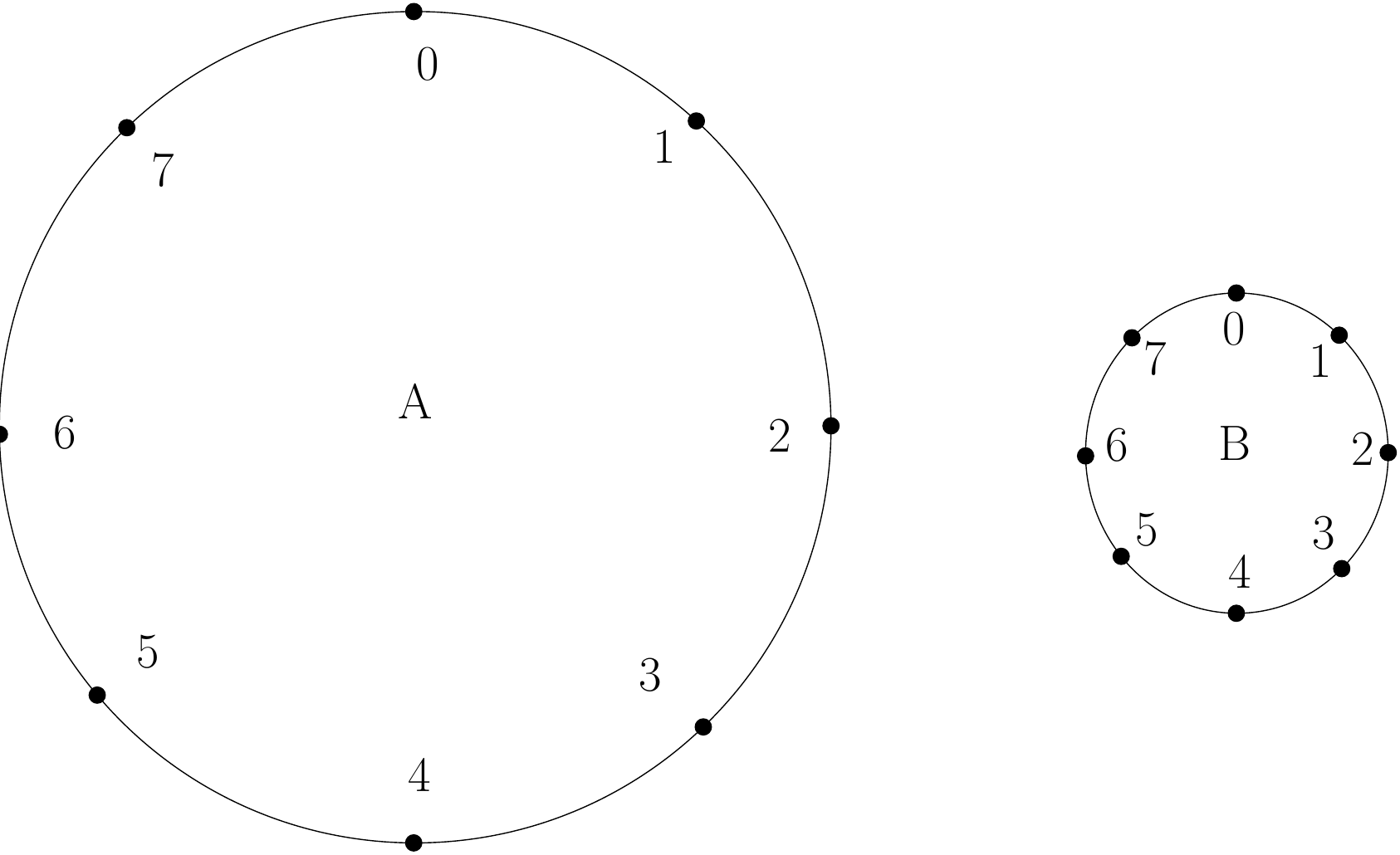}
 \caption{A conceptualization of the Composite code.
          $B$ moves clockwise until it reaches $0$,
          at which point $A$ moves clockwise one position also.}
 \label{fig:composite}
\end{figure}

During each generating step, the last $d'$ bits are considered
and moved to their
next state in the code generated by the rules of $B$, and checked to see if they
have reached their initial position,
The first $d$ bits are only incremented when the last
$d'$ bits cycle back to their initial
state - once for every $2^{d'}$ bit strings generated by the combined DAT.

Incrementing the last $d'$ bits requires $\le 6 \log d'$ bits to be read
on average and $1$ bit to be written.
Checking if the $d'$-code has reached its initial state
is equivalent to comparing it
against a fixed bit string (that is the code's initial state).  This is done
an equal number of times for all possible bit strings of dimension $d'$, and so
the average number of bits read is equal to the
expected position of the left-most $1$ in a random bit string of
dimension $d'$ which is equal to $\frac{2^{d'+1}-2-d'}{2^{d'}} \le 2$.

If we let $d' = \lceil\log r\rceil$, then the RPGC $B$ has dimension at least
$2$, from our restriction on $r$.  The average read cost of the
combined code to generate each bit string in the sequence becomes
no more than the average cost of the $d'$-code, plus the cost to check
if the code reached its initial state,
plus the cost of $d$-code divided by the length of the $d'$-code.

Therefore, the average number of bits read by the combined code is at most
\begin{align*}
	6\log d' + 2 + \frac{r}{2^{d'}}
	&= 6\log \lceil\log r\rceil + 2 + \frac{r}{2^{\lceil\log r\rceil}} \\
	&\le 6\log \lceil\log r\rceil + 3
	\text{ .}
\end{align*}

In the case that both $A$ and $B$ are incremented, the bits of $B$ are
equal to its initial state, so checking this requires reading all $d'$ bits.
This happens for all possible values in $A$, which requires reading $p$ bits
in the worst case.  Therefore, at some time, the new code will read
$p+d' = p + \lceil \log r \rceil$ bits.

The number of writes in the worst case, is the number of bits that must
change to update both the code contained in the first $d$ bits, and the code
contained in the last $d'$ bits.  Thus the number of
bits written per increment operation is at most
$\text{(bits written by $d$-code)} +
 \text{(bits written by $d'$-code)} \le w + 1$.
\end{proof}

\section{RPGC-Composite Code}
\label{subsec:rpgc-composite}

We are able to use the RPGC from Theorem \ref{thm:partcode} with our
Composite code from Lemma \ref{lem:composite}
to construct a new space-optimal
DAT that generates a code.
By applying Lemma \ref{lem:composite} to the RPGC, and then repeatedly
applying it $c-1$ more times to the resulting DAT,
we create a DAT that generates a code while reading on
average no more than $6 \log^{(2c-1)}d + 11$ bits,
and never writing more than $c$ bits to generate each bit string,
for any constant $c \ge 1$.

We use the following lemmas to prove Theorem \ref{thm:supercode}.

\begin{lemma}\label{lem:logrelation}
 Given $r \ge 16$, then $6\log(\log r+1) \le 6\log\log r + 2$.
\end{lemma}
\begin{proof}
Let $x = \log r$.  Then we want to prove that given $x \ge 4$,
$6\log(x+1) \le 6\log x + 2$.  It is equivalent to show that
$\log (1 + 1/x) \le \log(1+1/4) \le 1/3$.
Therefore $6\log(\log r+1) \le 6\log\log r + 2$.
\end{proof}

\begin{lemma}\label{lem:cdthing}
Let $c \ge 1$ and $d$ be integers such that $\log^{(c-1)}d \ge 2$.
Then $\log^{(c)}(d/2) \ge \log^{(c)}d - 1$.
\end{lemma}
\begin{proof}
The proof is by induction.  Let $c=1$.  Then $\log(d/2) = \log d - 1$.

Assume the claim is true for $c$, we will show that it is then true for $c+1$.
From our choice of $c$ and $d$, we know that
\begin{align*}
 \log^{(c)}d &\ge 2 \\
 2\log^{(c)}d - 2 &\ge \log^{(c)}d \\
 \log(2(\log^{(c)}d - 1)) &\ge \log(\log^{(c)}d) \\
 1 + \log(\log^{(c)}d - 1) &\ge \log^{(c+1)}d \\
 \log(\log^{(c)}d - 1) &\ge \log^{(c+1)}d - 1\\
 \log(\log^{(c)}(d/2)) &\ge \log^{(c+1)}d - 1 \text{\emph{ (from our inductive hypothesis)}}\\
 \log^{(c+1)}(d/2) &\ge \log^{(c+1)}d - 1
 \text{ ,}
\end{align*}
as required.
\end{proof}

\begin{lemma}\label{lem:mdrelation}
	Let $d$ and $c \ge 1$ be integers such that
$\log^{(2c+1)}d \ge 11$, and
let $m = d - \lceil\log(6\log^{(2c-1)}d+11)\rceil$.  Then
$\log^{(2c-1)}m \ge 11$.
\end{lemma}

\begin{proof}\label{pf:mdrelation}
Given that $\log^{(2c+1)}d \ge 11$, then we can say that
$\log\log d + 4 \le d/2$ because $d \ge 16$.  By our choice of $d$, we know that
$\log d \ge 11$ and thus $d \ge 2^{11} \ge 16$.  Using this, we can show that
$m \ge d/2$ as follows:
\begin{align*}
 m &=   d - \lceil\log(6\log^{(2c-1)}d+11)\rceil \\
   &\ge d - \log(6\log^{(2c-1)}d+11) - 1 \\
   &\ge d - \log(6\log^{(2c-1)}d+\log^{(2c-1)}d) - 1 \\
   &=   d - \log(7\log^{(2c-1)}d) - 1 \\
   &=   d - \log^{(2c)}d - \log 7 - 1 \\
   &\ge d - (\log^{(2c)}d + 4) \\
   &\ge d - (\log\log d + 4) \\
   &\ge d/2
   \text{ .}
\end{align*}

By our choice of $c$ and $d$ we know that
$\log^{(2c-1)}d = \log(\log^{(2c-2)}d) \ge 11$
and $\log^{(2c-2)}d \ge 2^{11} \ge 2$.
Then, since $m \ge d/2$, and with Lemma \ref{lem:cdthing}, it follows that
$\log^{(2c-1)}m \ge \log^{(2c-1)}(d/2) \ge \log^{(2c-1)}d - 1
 \ge 2^{2^{11}}-1 \ge 11$.
\end{proof}

\begin{thm}\label{thm:supercode}
	Given integers $d$ and $c \ge 1$, such that $\log^{(2c-1)}d \ge 11$.
There exists a DAT of dimension $d$ that generates a code of
length $L=2^d$, where generating the next bit
string requires reading on average no more than
$6\log^{(2c-1)}d+11$ bits and writing in the worst case at most $c$ bits.
\end{thm}

\begin{proof}\label{pf:supercode}
	The proof is by induction on $c$.  Let $c=1$.  In this case, the Recursive
Partition Gray Code in Section \ref{subsec:partition} satisfies the
requirements.  From our choice of $d$, the RPGC will have dimension at least
two.  It requires reading no more
than $6\log d \le 6\log^{(1)}d + 11$ bits on average, and
writes $1$ bit in each generating step.

	Then let $c \ge 1$, and assume that the theorem is true for $c$.  We
will show it is true for $c+1$.

Let $d$ be such that $\log^{(2c+1)}d \ge 11$, and
define $m = d - \lceil \log(6\log^{(2c-1)}d+11) \rceil$.
Then from Lemma \ref{lem:mdrelation} we know that $\log^{(2c-1)}m \ge 11$.
Thus we assume our initial claim holds true for $c$, and there
exists a DAT $M$ of dimension $m$ that requires reading on average no
more than $6\log^{(2c-1)}m + 11 \le 6\log^{(2c-1)}d+11$ bits,
and writing at most $c$ bits for each transition.

	We use the construction from Lemma \ref{lem:composite}
to perform the inductive step and construct a new DAT.
As input to Lemma \ref{lem:composite}, we use our DAT $M$,
which gives us $r=6\log^{(2c-1)}d+11 \ge 3$ and $w=c$.
This produces a new composite DAT of dimension
$m+\lceil\log r\rceil$ $=$ $m+\lceil \log(6\log^{(2c-1)}d+11) \rceil$ $=$ $d$
that reads no more than $6\log \lceil\log r\rceil + 3$ bits.
Note that from our choice of $c$ and $d$, $\log 7 \le \log 11 \le log^{(2c)}d$.
Then our new DAT will read on average no more than
\begin{align*}
 6\log \lceil\log r\rceil + 3
 &\le 6\log (\log r + 1) + 3 \\
 &\le 6\log\log r + 5 \text{ (from Lemma \ref{lem:logrelation})} \\
 &= 6\log\log(6\log^{(2c-1)}d+11) + 5 \\
 &\le 6\log\log(6\log^{(2c-1)}d + \log^{(2c-1)}d) + 5 \\
 &= 6\log\log(7\log^{(2c-1)}d) + 5 \\
 &= 6\log(\log 7 + \log^{(2c)}d) + 5 \\
 &\le 6\log(\log^{(2c)}d + \log^{(2c)}d) + 5 \\
 &= 6\log(2\log^{(2c)}d) + 5 \\
 &= 6\log 2 + 6\log^{(2c+1)}d) + 5 \\
 &= 6\log^{(2c+1)}d) + 11 \\
\end{align*}
bits, as required.
	
	From Lemma \ref{lem:composite}, in the worst case, our DAT writes at most
$w+1 = c+1$ bits to generate the next bit string.
These satisfy our initial claim, finishing the proof.

Thus, there exists a DAT for any $c \ge 1$, when $\log^{(2c-1)}d
\ge 11$, that generates all $2^d$ bit strings of a code of
dimension $d$, and requires reading on average
no more than $6\log^{(2c-1)}d+11$ bits, and writing at most $c$ bits to
generate each successive bit string from the previous string.
\end{proof}

\section{Reading a constant average number of bits}
\label{subsec:const-read}

We have shown in Section \ref{subsec:rpgc-composite} that it is possible to
construct a code that requires reading, on average, at most
$6\log^{(2c-1)}d+11$ bits, and never requires writing more than $c$ bits, for
any constant $c \ge 1$.  This allows reading a small number of bits while
writing at most a constant amount.

From Theorem \ref{thm:supercode}, by taking $c$ to be a function of $\log^*d$,
it immediately follows that we can create a
DAT that generates all $2^d$ bit strings of dimension $d$, for which each
generating step requires reading a constant number of
bits on average.  This is a trade off, as the DAT requires writing at
most $O(\log^*d)$ bits in the worst case, meaning
the code generated by this DAT is not considered
a quasi-Gray code.

\begin{corollary}\label{cor:logstar}
	There exists
	a space-optimal DAT of dimension $d$ which has the following properties
	for each generating step:

\begin{enumerate}
 \item For any $d > 2^{16}$, the DAT reads no more than
	$6 \cdot 2^{2^{16}} + 11$ bits on average, and writes no more than
	$\lfloor (\log^*d - 1)/2 \rfloor$ bits in the worst case.
 \item For any $d > 3 + 2^{17}$, the DAT reads no more than
	$100$ bits on average, and writes no more than
	$\lfloor (\log^*d + 1)/2 \rfloor$ bits in the worst case.
 \item For any $d > 10 + 2^{17}$, the DAT reads no more than
	$20$ bits on average, and writes no more than
	$\lfloor (\log^*d + 3)/2 \rfloor$ bits in the worst case.
 \item For any $d > 15 + 2^{17}$, the DAT reads no more than
	$17$ bits on average, and writes no more than
	$\lfloor (\log^*d + 5)/2 \rfloor$ bits in the worst case.
\end{enumerate}
\end{corollary}

\begin{proof}\label{pf:logstar}
Let $d > 2^{16}$, and $c = \lfloor (\log^*d-3)/2 \rfloor$.
Then $c \ge 1$ and $\log^{(2c-1)}d \ge 11$ and
it follows from Theorem \ref{thm:supercode} that there exists a
space-optimal DAT of dimension $d$, which has the following properties for each
generating step: In the average case, it requires reading no more than
$6\log^{(2\lfloor (\log^*d-3)/2 \rfloor-1)}d+11 \le
6\log^{(2 ((\log^*d-3)/2 -1/2)-1)}d+11 \le
6\log^{(\log^*d-5)}d+11 \le 6 \cdot 2^{2^{16}} + 11$ bits.
And in the worst case, it requires
writing at most $\lfloor (\log^*d - 1)/2 \rfloor$ bits.

Let $d > 3+2^{17}$, and $m = d-(3+2^{16}) > 2^{16}$. From our previous statement,
we can construct a DAT of dimension $m$, which
requires reading at most $6 \cdot 2^{2^{16}} + 11 \le 2^{3+2^{16}}$
bits on average, and writing no more than
$\lfloor (\log^*m-1)/2 \rfloor \le \lfloor (\log^*d-1)/2 \rfloor$
bits in the worst case.  Use this DAT
with Lemma \ref{lem:composite}, setting
$r=2^{3+2^{16}}$ and $w=\lfloor (\log^*d-1)/2 \rfloor$.
Then there exists a DAT of dimension
$m+\lceil\log r\rceil = m+3+2^{16}=d$, that requires reading on average
no more than $6\log\lceil\log r\rceil+3 \le 100$ bits to
generate the next bit string in the code, and writes no more
than $\lfloor (\log^*d + 1)/2 \rfloor$ bits in the worst case.

Let $d > 10 + 2^{17}$, and $m = d-7 > 3+2^{17}$.
Use the DAT from our previous statement with Lemma \ref{lem:composite},
setting $r=100$ and $w=\lfloor (\log^*d+1)/2 \rfloor$.
Then there exists a DAT of dimension
$m + \lceil\log r\rceil = m+7 = d$, which requires reading on average
no more than $6\log\lceil\log r\rceil+3 \le 20$ bits to generate the next
bit string in the code, and writes no more
than $\lfloor (\log^*d + 3)/2 \rfloor$ bits in the worst case.

Let $d > 15 + 2^{17}$, and $m = d-5 > 10 + 2^{17}$.
Use the DAT from our previous statement with Lemma \ref{lem:composite},
setting $r=20$ and $w=\lfloor (\log^*d+3)/2 \rfloor$.
Then there exists a DAT of dimension
$m + \lceil\log r\rceil = m+5 = d$, which requires reading on average
no more than $6\log\lceil\log r\rceil+3 \le 17$ bits to generate the next
bit string in the code, and writes no more
than $\lfloor (\log^*d + 5)/2 \rfloor$ bits in the worst case.
\end{proof}

\section{Lazy counters}
\label{subsec:lazycounters}

A lazy counter is a structure for generating a sequence of bit strings.
In the first $n$ bits, it counts through the standard binary representations of $0$
to $2^n-1$.  However, this can require updating up to $n$ bits, so an additional
data structure is added to slow down these updates, making it so that each
successive state requires fewer bit changes to be reached.
We present a few known lazy counters, and then improve upon them,
using our results to generate quasi-Gray codes.

\subsection{LazyIncrement}

Frandsen \textit{et al.} \cite{256309} describe a lazy counter of dimension $d$
that reads and writes at most $O(\log d)$
bits for an increment operation.  The algorithm uses $d = n + \log n$ bits,
where the first $n$ are referred to as $b$, and the last $\log n$ are referred to
as $i$.  A state in this counter is the concatenation of $b$ and $i$,
thus each state is a bit string of dimension $d$.  In the
initial state, all bits in $b$ and $i$ are set to $0$.  The counter then
moves through $2^{n+1}-2$ states before cycling back to the initial state,
generating a cyclic code.

The bits of $b$ move through the standard binary numbers.
However, moving from one such number to the next may require writing as many
as $n$ bits.  The value in $i$ is a pointer into $b$.
For a standard binary encoding, the algorithm to move from one
number to the next is as follows: starting at the right-most
(least significant)
bit, for each $1$ bit, flip it to a $0$ and move left.  When a $0$ bit is
found, flip it to a $1$ and stop.  Thus the number of bit flips required to
reach the next standard binary number is equal to one plus the position of the
right-most $0$.  This counter simply uses $i$ as a pointer into $b$ such that
it can flip a single $1$ to a $0$ each increment step until $i$ points to a $0$,
at which point it flips the $0$ to a $1$, resets $i$ to $0$, and $b$ has then
reached the next standard binary number.  The algorithm is stated
as follows:

\begin{algorithm}[H]
	\caption{LazyIncrement \cite{256309}}
	\label{proc:lazyincrement}

	\KwIn{$b[]$: an array of $n$ bits; $i$: an integer of $\log n$ bits}
	\eIf{$b[i] = 1$}{
		$b[i] \leftarrow 0$\;
		$i \leftarrow i+1$\;
	}{
		$b[i] \leftarrow 1$\;
		$i \leftarrow 0$\;
	}
\end{algorithm}

\begin{lemma}\label{lem:lazyincrement} \cite{256309}
	Let $n \ge 2$ be a power of two.
	There exists a DAT of dimension $d = n + \log n$, using
	the \texttt{LazyIncrement} algorithm, that generates
	$2^{n+1}-2$ of a possible $n2^n$ bit strings,
	where in the limit $n \to \infty$ the
	space efficiency drops to $0$.  The DAT reads and writes in the
	worst case $\log n + 1$ bits to generate each successive bit string,
	and on average reads and writes $3$ bits.
\end{lemma}

\begin{proof}
The maximum
number of bit strings generated by this counter is $2^d = n 2^n$,
however it actually generates significantly less.

The bits of $b$ move through each standard binary number, but with
additional states in between, such that each state in differs by a single bit
in $b$.  Thus, the number of states between two of these numbers is equal
to the number of the bits that need to change, and this is equal to the
distance to the right-most $0$ bit in $b$.

The right-most bit in $b$ is $0$ for half of the standard binary numbers.  For
the other half, there is a $0$ to the left of it half of the time.  If the
right-most bit is position $1$, then bit $j$ contains the right-most $0$
exactly $2^n/2^j$ times, over all standard binary numbers which
fit in $n$ bits.  Meanwhile, the bits in $i$ make it
possible to have these transitional states, but don't provide any
additional states beyond them.
Thus we can count the number of total states in the
counter.  Bit $j$ is the right-most $0$ in $b$
for $2^n/2^j$ numbers, and
each time it is, it requires $j$ bit flips to reach the next number.
Additionally, when there are no $0$ bits in $b$ at all, $n$ bit flips are
required to reach the next number, that returns $b$ back to all zeros.
Let $G(d)$ be the number of bit strings generated by \texttt{LazyIncrement}, for
dimension $d = n + \log n$.  Then the number of generated bit strings is
$G(d) = \sum_{j=1}^{n} j \frac{2^n}{2^j} + n
	= 2^n \sum_{j=1}^{n} \frac{j}{2^j} + n
	= 2^{n+1} - 2$.

The space efficiency of this counter, or the ratio of bit strings generated to
the number of possible strings is
$\frac{G(d)}{2^{d}} = \frac{2^{n+1}-2}{2^{n+\log n}} = \frac{2^n-1}{n 2^{n-1}}$.
Since $\lim_{n\to\infty} \frac{2^n-1}{n 2^{n-1}} = 0$,
the counter is non-space-optimal, and the space efficiency of
the counter grows worse as its dimension grows larger.

To generate each successive bit string, the worst-case number of bits
read or written by this counter is the cost to
increment or reset $i$,
which is an integer of dimension $\log n$, plus the single bit which changes
in $b$.  Thus the counter never reads nor writes more than $\log n + 1$ bits.
On average, incrementing an integer (using the
standard binary representation) requires reading and writing at
most $2$ bits, while only a single bit is ever read or written in $b$ at a time.
Thus the average cost to generate a bit string in this counter is at most $3$.

The \texttt{LazyIncrement} algorithm presented above can be constructed as DAT.
First construct a DAT that reads the value of $i$.
This tree reads all $\log n$ bits of $i$ and has $n$ leaves, one for each
value of $i$.  The leaf that represents $i = x$ will then read bit $x$ of
$b$.  If the bit was a $0$, it is changed to a $1$, and all of the bits in
$i$ are set to $0$.  If the bit was a $1$, then the bit is changed to a $0$
and $i$ is changed to represent $x+1$.
\end{proof}

\subsection{SpinIncrement}

An observation by Brodal \cite{Brodal} (unpublished) leads to
a dramatic improvement in space efficiency over the previous
algorithm by adding a single bit to the counter.  This extra bit allows for
the $\log n$ bits in $i$ to spin through all their possible values, thus making
better use of the bits and generating more bit strings with them.
The variables $b$ and $i$ are unchanged from the counter in
Lemma \ref{lem:lazyincrement}, and $k$ is a single bit, making the counter have
dimension $d = n + \log n + 1$.  The algorithm is as follows.

\begin{algorithm}[H]
	\caption{SpinIncrement \cite{Brodal}}
	\label{proc:spinincrement}

	\KwIn{$b[]$: an array of $n$ bits; $i$: an integer of $\log n$ bits;
	      $k$: a single bit}
	\eIf{$k = 0$}{
		$i \leftarrow i+1$  \tcp*[l]{spin $i$}
		\If{$i = 0$}{
			$k \leftarrow 1$  \tcp*[l]{the value of $i$ has rolled over}
		}
	}{
		LazyIncrement($b[]$, $i$)  \tcp*[l]{really increment the counter}
	    \If{$i = 0$}{
	      $k \leftarrow 0$\;
	    }
	}
\end{algorithm}

\begin{lemma}\label{lem:spinincrement} \cite{Brodal}
	Let $n \ge 2$ be a power of two.
	There exists a DAT of dimension $d = n + \log n$, using
	the \texttt{SpinIncrement} algorithm, that generates
	$(n+1)(2^n-1)$ of a possible $2 n 2^n$ bit strings,
	where in the limit $n \to \infty$ the
	space efficiency converges to $1/2$.
	The DAT reads and writes in the
	worst case $\log n + 2$ bits to generate each successive bit string,
	and on average reads at most $4$ bits.
\end{lemma}

\begin{proof}
This counter spins through $i$ every time that $b$ becomes equal to its next
value in the standard binary representation, for which there are $2^n$ such
cases.
Spinning through all values for $i$ adds exactly $2^{\log n} = n$ states.
Thus the total number of states for this improved lazy counter is $n2^n
+ 2^{n+1}-2 - 2^n = n2^n + 2^n-2 = (n+1)2^n-2$.

The space efficiency of this improved counter is
$\frac{(n+1)2^n-2}{2 n 2^n}$.  In the limit $n \to \infty$ the
space efficiency converges to $1/2$.
That is, when the counter has large dimension $d$, approximately half of the
$2^d$ possible bit strings are generated.

The worst-case number of bits read or written by this counter is one more than
the counter in Lemma \ref{lem:lazyincrement}.
The only cases added are where $i$
and $k$ are changed, which together do not exceed that bound.  However, in the
case where $b$ and $i$ are changed, $k$ may also be read and changed, so the
counter may read and write at most $\log n + 2$ bits to generate the next bit
string.
Checking the value of $k$ requires reading a single bit, incrementing $i$
requires on average to read and write at most $2$ bits.  Comparing $i$ for
equality to the fixed bit string $000...0$
can be done for free after incrementing $i$ in the DAT model.
The \texttt{LazyIncrement} algorithm reads at most $3$ bits on average, after
$k$ has been read, and is run an average of $2$ times with $n$ increments
in between, contributing at most $4$ to the average. From these
observations, it follows that the average number of bits read and written
by this counter, to generate the next bit string, does not exceed $4$.

The \texttt{SpinIncrement} algorithm can be constructed as DAT
using the DAT from Lemma \ref{lem:lazyincrement}.
Add a new root node that reads $k$.
When $k = 0$, go to its left child.  The
right child is a subtree that reads all the bits of $i$.  For all leaf
nodes of this subtree, $i$ is incremented.  And when $i$ was equal to
$n-1$, $k$ is also set to $0$.
When $k = 1$, go to its right child,
which is the root of a subtree mostly identical to the
DAT for \texttt{LazyIncrement}, with an extra rule in the leaf where $i$
is set to $000...0$, that sets $k$ to $0$ as well.
\end{proof}

\subsection{DoubleSpinIncrement}

By generalizing the dimension of $k$, we are able to make the counter even more
space efficient while keeping its $O(\log n) = O(\log d)$ worst-case
bound for bits written and read.
Let $k$ be a bit array of dimension $g \ge 1$.
Then for a counter of dimension $d = n + \log n + g$, the new algorithm
is as follows.

\begin{algorithm}[H]
	\caption{DoubleSpinIncrement}
	\label{proc:doublespin}

	\KwIn{$b[]$: an array of $n$ bits; $i$: an integer of $\log n$ bits;
	      $k$: an integer of $g$ bits}
	\eIf{$k < 2^g - 1$}{									\nllabel{line:ds1}
		$i \leftarrow i+1$
			\tcp*[l]{happens in $(2^g-1)/2^g$ of the cases}	\nllabel{line:ds2}
		\If{$i = 0$}{										\nllabel{line:ds3}
			$k \leftarrow k+1$\;
		}
	}{
	    LazyIncrement($b[]$, $i$)  \tcp*[l]{do a real increment}
	    \If{$i = 0$}{
	      $k \leftarrow 0$\;
	    }
	}
\end{algorithm}

\begin{thm}\label{thm:doublespinincrement}
	Let $n \ge 2$ be a power of two, and $g \ge 1$ be an integer.
	There exists a DAT of dimension $d = n + \log n + g$ with
	space efficiency $1-O(2^{-g})$.
	The DAT, using the \texttt{DoubleSpinIncrement} algorithm,
	reads and writes in the
	worst case $g + \log n + 1$ bits to generate each successive bit string,
	and on average reads and writes $O(1)$ bits.
\end{thm}

\begin{proof}
This counter generates $2^g-1$ states for each time it
spins through the possible values of $i$.  Thus the number of bit strings
generated is $n2^n(2^g-1) + 2^n-2 = n2^n2^g-(n-1)2^n-2$.
Given the dimension of the counter,
the possible number of bit strings generated is $n 2^n 2^g$.  This gives a
space efficiency of $1-O(2^{-g})$.
When $g = 1$, we have exactly the same counter as given by
Lemma \ref{lem:spinincrement}, and when $g$ is an increasing function of $n$,
we produce a counter with space efficiency arbitrarily close to one.

In the worst case, this counter reads and writes every bit in $i$ and $k$, and
a single bit in $b$, thus $g + \log n+1$ bits.
On average, the counter now reads and writes $O(1)$ bits.  This follows from
a similar argument to that made for Lemma \ref{lem:spinincrement}, where each
line modified in the algorithm still reads on average $O(1)$ bits.

%

The \texttt{DoubleSpinIncrement} can also be constructed as DAT
by building on the DAT from Lemma \ref{lem:spinincrement}.
Replace the root node with a subtree that
reads all the bits of $k$ and has $2^g$ leaf nodes.  For leaf nodes that
read a value in $k$ less than $2^g-1$, the leaf node becomes the root of a
subtree similar to the right child of the root node in the DAT for
\texttt{SpinIncrement}.  These subtrees are modified in that $k$ is
incremented instead of set to $0$ when $i$ was equal to $n-1$.  The one
leaf node where $k = 2^g-1$ becomes the root of a subtree identical to the
left child of the root node in the DAT for \texttt{SpinIncrement}.
\end{proof}

\begin{corollary}\label{cor:doublespint}
	Let $n \ge 2$ be a power of two, and 
	$g = t \log n$ be an integer, for $t > 0$.
	There exists a DAT of dimension $d = n + (t+1)\log n$
	with space efficiency $1-O(n^{-t})$.
	The DAT, using the \texttt{DoubleSpinIncrement} algorithm,
	reads and writes in the
	worst case $(t+1) \log n + 1$ bits to generate each successive bit string,
	and on average reads and writes $O(1)$ bits.
\end{corollary}
\begin{proof}
	This follows from Theorem \ref{thm:doublespinincrement} by
substituting $g = t \log n$.
\end{proof}

\subsection{WineIncrement}

Rahman and Munro \cite{Rahman08} present a counter that reads at most
$\log n + 4$ bits and writes at most $4$ bits to perform an increment or
decrement operation.  The counter uses $n+1$ bits to count through $2^n$
states, and has space efficiency $2^n/2^{n+1} = 1/2$.
Compared to \texttt{DoubleSpinIncrement},
their counter writes fewer bits per generating step,
but is less space efficient.
By modifying our
lazy counter to use Gray codes internally, the worst-case
number of bits read remains asymptotically equivalent to the counter by
Rahman and Munro, and the average number of bits we read increases.
We are able to write a smaller constant number of bits per increment and
retain a space efficiency arbitrarily close to $1$.

We modify our counter in Theorem \ref{thm:doublespinincrement}
to make $i$ and $k$ hold a cyclic Gray
code instead of a standard binary number.  The BRGC is a suitable
Gray code for this purpose, so we will use it.  Given a function
\texttt{next}($j$) that takes a bit string $j$ of rank $r$ in the BRGC and returns the
bit string of rank $r+1$, and a function \texttt{rank}($j$) that returns the rank value
of the bit string $j$ in the BRGC.  The following algorithm provides a lazy
counter of dimension $d = n + \log n + g$, where $g \ge 1$,
that writes at most $3$ bits, and
reads at most $g + \log n + 1$ bits to generate the next state,
and is space-optimal in the limit $n \to \infty$.

\begin{algorithm}[H]
	\caption{WineIncrement\protect\footnotemark}
	\label{proc:wine}

	\KwIn{$b[]$: an array of $n$ bits; $i$: a Gray code of $\log n$ bits;
	      $k$: a Gray code of $g$ bits}
	\eIf{$k \ne 100...00$}{									\nllabel{line:inc1}
		$i \leftarrow$ \texttt{next}($i$)
			\tcp*[l]{happens in $(2^g-1)/2^g$ of the cases}
															\nllabel{line:inc2}
		\If{i = 0}{											\nllabel{line:inc3}
			$k \leftarrow$ \texttt{next}($k$)\;						\nllabel{line:inc4}
		}													\nllabel{line:inc5}
	}{														\nllabel{line:inc6}
		\tcp{do a real increment}							\nllabel{line:inc7}
		\eIf{$b[$\emph{\texttt{rank}($i$)}$] = 1$}{				\nllabel{line:inc8}
			$b[$\texttt{rank}($i$)$] \leftarrow 0$\;		\nllabel{line:inc9}
			$i \leftarrow$ \texttt{next}($i$)\;						\nllabel{line:inc10}
			\If{$i = 0$}{									\nllabel{line:inc11}
				$k \leftarrow 0$
					\tcp*[l]{wraps around to the initial state}
															\nllabel{line:inc12}
			}												\nllabel{line:inc13}
		}{													\nllabel{line:inc14}
			$b[$\texttt{rank}($i$)$] \leftarrow 1$\;		\nllabel{line:inc15}
			$k \leftarrow 0$
				\tcp*[l]{resets $k$ to $0$}					\nllabel{line:inc16}
		}													\nllabel{line:inc17}
	}														\nllabel{line:inc18}
\end{algorithm}

\footnotetext{The name \texttt{WineIncrement} comes from the
Caribbean dance known as Wineing,
a dance that is centered on rotating your hips with the music.
The dance is pervasive in Caribbean culture, and has been popularized elsewhere
through songs and music
videos such as Alison Hinds' ``Roll It Gal'', Destra Garcia's ``I Dare You'',
and Fay-Ann Lyons' ``Start Wineing''.}

\begin{thm}\label{thm:winebrgcincrement}
	Let $n \ge 2$ be a power of two, and $g \ge 1$ be an integer.
	There exists a DAT of dimension $d = n + \log n + g$ with
	space efficiency $1-O(2^{-g})$.
	The DAT, using the \texttt{WineIncrement} algorithm,
	reads in the worst case $g + \log n + 1$ bits and writes in the
	worst case $3$ bits to generate each successive bit string.
\end{thm}

\begin{proof}
\begin{sloppypar}
The number of states used by this counter would be the same as
the \texttt{DoubleSpinIncrement} algorithm, except that we are unable to
reset $i$ to $0$ when a bit in
$b$ is flipped to $1$ without changing up to $\log n$ bits in $i$.
Instead, we leave $i$ as it is and observe that it
does not significantly reduce the space efficiency of the counter.
\end{sloppypar}

Note that when a bit in $b$ is flipped to a $1$:
\begin{itemize}
	\item $rank(i)$ points to the index of the right-most $1$ in $b$,
counting from the least-significant bit.
	\item a bit in $b$ flips to $1$ if and only if the bit sequence $b$ is
entering the next state of the standard binary number encoding.  That
is, $b$ counts from $0$ thru $2^n-1$, and back to $0$,
as a standard binary number, with extra states in between.
The extra states all come from steps where a bit in $b$ flips to $0$.
\end{itemize}

Based on these observations we can sum up all the values of $i$ that occur
when a bit in $b$ flips to $1$ as the positions of the right-most
$1$ in all possible bit strings of dimension $n$, which
is $2^n\sum_{i=1}^n\frac{i}{2^i} = 2^{n+1}-n-2$.
When $i$ has a value greater than $0$, the number of states
lost compared to \texttt{DoubleSpinIncrement} is exactly the difference between
$i$ and $0$.  Thus, over all increment steps, this summation describes the
total number of states lost compared to our third lazy counter.  Therefore
the total number of states used is
$n2^n2^g-(n-1)2^n-2 - 2^{n+1} + n + 2 = n2^n2^g-(n+1)2^n + n$.
The number of bits, and thus, the number of possible states, is unchanged
from the \texttt{DoubleSpinIncrement} algorithm.  There are $2^{n+\log n+g}$
possible states,
and the space efficiency of this counter as $n$ grows large converges to
\[
 \lim_{n\to\infty}
		\frac{n2^n2^g-(n+1)2^n + n}{n2^n2^g}
 = 1 - \lim_{n\to\infty}
		    \frac{(n+1)2^n - n}{n 2^n 2^g}
 = 1 - O(2^{-g})
\text{ .}
\]

The average number of bits read for each line of the algorithm is $O(1)$,
with the exception of the lines which increment $i$ and $k$.
When they are executed, these lines read on average $\log n$ and $g$ bits
respectively, so the total average number of bits read is
at most $\log n + g + 1$.

The counter writes at most one bit in each of $b$, $i$, and $k$, and thus writes
in the worst case $3$ bits per increment operation.
\end{proof}

\begin{corollary}\label{cor:winebrgct}
	Let $n \ge 2$ be a power of two, and 
	$t \log n$ be an integer, for $t > 0$.
	There exists a DAT of dimension $d = n + (t+1)\log n$
	with space efficiency $1-O(n^{-t})$.
	The DAT, using the \texttt{WineIncrement} algorithm,
	reads in the worst case $(t+1) \log n + 1$ bits and writes in the
	worst case $3$ bits to generate each successive bit string.
\end{corollary}
\begin{proof}
	This follows from Theorem \ref{thm:winebrgcincrement} by
substituting $g = t \log n$.
\end{proof}

From Corollary \ref{cor:winebrgct}, we get a counter that is more
space efficient than previously known counters, with a constant number
of bits written for each increment.
But while the counter reads at most $(t+1) \log n + 1$ bits
in the worst case, its average number of bits read is also $O(\log n)$.
Using the quasi-Gray
code counter from Theorem \ref{thm:supercode}, we are able to bring the
average number of bits read down as well.  The worst case number of bits
read remains $(t+1) \log n + 1$, but on average, we only read at most
$12\log^{(2c)}n + O(1)$ bits, for any $c \ge 1$.

The algorithm does not need to change from its fourth iteration for these
modifications.  We simply make $i$ a quasi-Gray code from Theorem
\ref{thm:supercode} of dimension $\log n$ and $k$ a similar quasi-Gray code
of dimension $g$.

\begin{thm}\label{thm:wineincrement}
	Let $n$ be a power of two such that $\log^{(2c)}n \ge 11$ and
	$g$ be an integer such that $\log^{(2c-1)}g \ge 11$.
	Then for any $c \ge 1$,
	there exists a DAT
	of dimension $d = n + \log n + g$ bits,
	using the \texttt{WineIncrement} algorithm,
	with space efficiency $1-O(2^{-g})$, that
	reads in the worst case $g + \log n + 1$ bits, writes in the
	worst case $2c+1$ bits,
	and reads on average no more than $12\log^{(2c)}n + O(1)$ bits.
\end{thm}

\begin{proof}
We determine the average number of bits read in the previous algorithm for
each line of the \texttt{WineIncrement} function.
Line \ref{line:inc1} becomes \texttt{true} each time $b$ reaches a
standard binary number representation, and there are $2^n$ such numbers.
Each time the line becomes \texttt{true}, it stays so while $i$ and $k$
spin through their values together.  During this time, the
value of $k$ changes every $n$ steps, and we see all possible values of $k$
except its maximally ranked state $1000...0$.
In the \texttt{DoubleSpinIncrement} algorithm, this would account for
$n2^n(2^g-1)$ states, but all states lost due to $i$ happen during this stage,
and so it accounts instead for
$n2^n(2^g-1) - 2^{n+1}+n+2 = n2^n2^g-(n+2)2^n+n+2$ states.

If we saw each value of $k$ once, the number of bits read to test
if the string is equal to $1000...0$ would be
$\sum_{i=1}^{g}i\cdot2^g/2^i = 2^{g+1} - g - 2$.  Instead, we see each
value in $k$ for $n$ consecutive steps, except for state $1000...0$.
Therefore, the number of bits read between states where $k=1000...0$ is
$n(2^{g+1} - g - 2 - g2^g/2^g) = n2^{g+1}-2ng-2n$.

During the remaining $2^n - 2$ states, $k=1000...0$, and line \ref{line:inc1}
will read all $g$ bits each time it is executed, reading a total of $g2^n-2g$
bits.
Thus the the average number of bits read by line \ref{line:inc1} is
\begin{eqnarray*}
	R_{\ref{line:inc1}}
        &=& \frac{n2^{g+1}-2ng-2n + g2^n - 2g}
	              {n2^n2^g-(n+1)2^n + n} \\
        &\le& \frac{n2^{g+1}+g2^n}{n2^{n+g}-(n+1)2^n+n} \\ 
        &=& \frac{1}{2^{n-1}}\frac{n2^{n+g}}{n2^{n+g}-(n+1)2^n+n}
           + \frac{g}{n2^g}\frac{n2^{n+g}}{n2^{n+g}-(n+1)2^n+n} \\
        &=& \frac{1}{2^{n-1}}\frac{n2^{n+g}-(n+1)2^n+n+(n+1)2^n-n}{n2^{n+g}-(n+1)2^n+n} \\
          && +\ \frac{g}{n2^g}\frac{n2^{n+g}-(n+1)2^n+n+(n+1)2^n-n}{n2^{n+g}-(n+1)2^n+n} \\
        &\le& \frac{1}{2^{n-1}}\left(1 + \frac{(n+1)2^n}{n2^{n+g}-(n+1)2^n+n}\right)
           + \frac{g}{n2^g}\left(1 + \frac{(n+1)2^n}{n2^{n+g}-(n+1)2^n+n}\right) \\
        &\le& 2 + \frac{2(n+1)}{n2^{n+g}-(n+1)2^n+n}
           + \frac{2^n}{n2^{n+g}-(n+1)2^n+n} \\
        &=& O(1)
	\text{ ,}
\end{eqnarray*}
since each of the two fractions are at most one when $n \ge 2$.

Line \ref{line:inc2} increments our counter from Theorem \ref{thm:supercode},
and reads on average
$R_{\ref{line:inc2}} \le 6\log^{(2c-1)}(\log n) + 11 = 6\log^{(2c)}n + 11$ bits.

Line \ref{line:inc3} checks if $i$ is equal to a specific value.  This
check is done
consecutively over all values of $i$.  Thus half the time, only one bit
needs to be checked, a quarter of the time, two bits, and so on.  The
average number of bits read for this line is then
$\sum_{i=1}^{\log n}i\frac{n}{2^i}$, each time it is reached.  Once the check
in line \ref{line:inc1} returns false, it does so once for each value of $k$
other than $10..00$.  Thus this line is executed $(2^g-1)(2^n-1)$ times.
The average number of bits read by line \ref{line:inc3} is
\begin{align*}
	R_{\ref{line:inc3}}
	    &\le (\sum_{i=1}^{\log n}i\frac{n}{2^i})
	          \frac{2^g (2^n-1)}
	               {n(2^g-1)(2^n-1) + 2^n - 2^{n+1} + n + 1} \\
	    &\le \frac{2n 2^g (2^n-1)}
	              {n(2^g-1)(2^n-1) - 2^n + n + 1} \\
        &= \frac{n 2^{g+1} (2^n-1)}
                {n (2^g-1) (2^n-1) - 2^n} \\
        &= \frac{2^{g+1}}
                {(2^g-1) - \frac{2^n}{n (2^n-1)}} \\
        &\le \frac{2^{g+1}}{(2^g-1)-1} \\
        &= 2\frac{(2^g-2)+2}{(2^g-2)} \\
        &= 2 + \frac{4}{(2^g-1)} \\
        &= O(1)
	\text{ .}
\end{align*}

Line \ref{line:inc4} is similar to line \ref{line:inc2}, but with a
slightly larger counter.  It
reads on average
$R_{\ref{line:inc4}} \le 6\log^{(2c-1)}(t\log n) + 11$
$= 6\log^{(2c-1)}t + 6\log^{(2c)}n + 11$ bits, for $c \ge 1$.

Line \ref{line:inc8} reads a single bit in $b$, and must read all $\log n$
bits in $i$
in order to determine its rank.  The line of code is executed $2^{n+1}-2$
times over the entire sequence of transitions for the counter as a whole. 
Thus the average number of bits read is
\begin{align*}
    R_{\ref{line:inc8}}
        &= \frac{(1+\log n)(2^{n+1} - 2)}
                {n(2^g-1)(2^n-1) + 2^n - 2^{n+1} + n + 1} \\
        &= \frac{2(2^n-1) + 2(2^n-1)\log n}
                {n(2^g-1)(2^n-1) - 2^n + n + 1} \\
        &\le \frac{2(2^n-1) + 2(2^n-1)\log n}
                  {n(2^g-1)(2^n-1) - 2^n} \\
        &= \frac{2 + 2\log n}
                {n(2^g-1) - 1 - \frac{1}{2^n-1}} \\
        &\le \frac{2 + 2\log n}
                  {n(2^g-1) - 2} \\
        &\le 2 + \frac{2\log n - 2n (2^g-1)}
                      {n(2^g-1)} \\
        &\le 2 + \frac{2\log n}{n(2^g-1)} \\
        &= O(1)
	\text{ .}
\end{align*}

Line \ref{line:inc9} reads all $\log n$ bits of $i$, similarly to
line \ref{line:inc8}, and is
executed no more times than is line \ref{line:inc8}.  Thus the average
number of bits
read for line \ref{line:inc9} is $R_{\ref{line:inc9}} \le O(1)$.

Line \ref{line:inc10} has the average cost as line \ref{line:inc2} per
execution, and is executed fewer times over all generated states,
thus $R_{\ref{line:inc10}} \le R_{\ref{line:inc2}}$ bits.

Line \ref{line:inc11} has an average cost per execution that is the same
as line \ref{line:inc3},
but is executed fewer times, at most $2^{n+1}-2$ times.  Thus its average
cost $R_{\ref{line:inc11}} \le O(1)$.

Line \ref{line:inc12} does not need to read any bits, as the
state of $k$ is already known from line \ref{line:inc1}.

Line \ref{line:inc15} is similar to line \ref{line:inc9}, and does
not read more bits on average
than line \ref{line:inc8}.  Its average cost is
$R_{\ref{line:inc15}} \le \ref{line:inc8}$.

Line \ref{line:inc16}, like line \ref{line:inc12}, does not need to read
any bits.

The total average number of bits read can be determined by the summation
$\sum_{i=1}^{\ref{line:inc16}}R_i$.  It is clear that this average is at most
$12\log^{(2c)}n + O(1)$.  The number of bits written in the worst case remains
constant, though it grows slightly to $2c+1$.  Thus with $c=1$,
we can read on average $O(\log\log n)$ bits, write at most $3$,
and read in the worst case $g + \log n + 1$ bits.
This is accomplished while keeping the space
efficiency arbitrarily close to $1$, as in the previous counter.

The \texttt{WineIncrement} algorithm can be constructed as a DAT as well,
as long as the quasi-Gray codes used in $i$ and $k$ can be generated with a DAT.
This is done by using the DAT structures for $i$ and $k$ inside the tree from
Theorem \ref{thm:doublespinincrement}.
For this DAT, we are using a quasi-Gray code instead of standard
integers for our counting variables.  While this changes the description of
the resulting DAT, the actual structure does not change dramatically. 
First, replace the root node with a subtree $T_k$ that reads all the bits
of $k$ and has $n^c$ leaf nodes, labeled from $1$ to $n^c$.  Let leaf $j$
be reached when $k$ has rank $j$ in its quasi-Gray code.
The highest rank leaf will
read the last bit string from the code in $k$,
and it becomes the root of another
subtree, $T_{i_{n^c}}$.  This subtree corresponds to the main \texttt{else}
clause of the algorithm.  The subtree reads $i$, and has $n$ leaves, one
for each possible state of $i$.  Again, let leaf $j$ be reached when $i$
has rank $j$ in its quasi-Gray code.
Then, in the leaf node $j$ of $T_{i_{n^c}}$, the
$j$\emph{-th} bit of $b$ is read.  If the bit was a $0$, it is changed to a $1$,
and $k$ is moved ahead one state to $0$.  If the bit was a $1$, then the
bit is changed to a $0$ and $i$ is changed to represent the state of rank
$j+1$ in its quasi-Gray code, and lastly if $j$ is $n-1$, that is
this leaf represents
the highest rank for the quasi-Gray code in $i$, additionally
move $k$ ahead one
state to return it to $0$.

Other leaf nodes of $T_k$ also become roots of subtrees $T_{i_u}$ for $1
\le u < n^c$, where $T_{i_u}$ is rooted at the $u$\emph{-th} child of $T_k$, and
is reached when $k$ has rank $u$ in its quasi-Gray code.
These subtrees represent the
many cases where the main \texttt{if} clause is true in the algorithm.  The
subtrees each read all of $i$ and have $n$ leaf nodes, one for each
possible state of $i$.  Each leaf node has a rule to modify $i$ such that
its rank in its quasi-Gray code increases one.  Additionally the $n$\emph{-th}
leaf of each subtree $T_{i_u}$, has a rule to modify $k$ to represent
the state of rank $u+1$ in its quasi-Gray code.
\end{proof}

\begin{corollary}\label{cor:winet}
	Let $n$ be a power of two
	such that $\log^{(2c)}n \ge 11$ and
	$t \log n$ be an integer, for $t > 0$
	such that $\log^{(2c-1)}(t \log n) \ge 11$.
	Then for any $c \ge 1$,
	there exists a DAT
	of dimension $d = n + (t+1)\log n$ bits,
	using the \texttt{WineIncrement} algorithm,
	with space efficiency $1-O(n^{-t})$, that
	reads in the worst case $(t+1) \log n + 1$ bits, writes in the
	worst case $2c+1$ bits,
	and reads on average no more than $12\log^{(2c)}n + O(1)$ bits.
\end{corollary}
\begin{proof}
	This follows from Theorem \ref{thm:wineincrement} by
substituting $g = t \log n$.
\end{proof}

\begin{corollary}\label{cor:winelogstar}
	Let $n > 2^{2^{16}}$ be a power of two, and
	$t \log n$ be an integer, for
	$t > 0$ such that $\log^{(2c-1)}(t \log n) \ge 11$.
	Then there exists a DAT
	of dimension $d = n + (t+1)\log n$ bits,
	with space efficiency $1-O(n^{-t})$, that
	reads in the worst case $(t+1) \log n + 1$ bits, writes in the
	worst case $log^*n-3$ bits,
	and reads on average $O(1)$ bits.
\end{corollary}
\begin{proof}
	Let $c=\lfloor(\log^*n-4)/2\rfloor$.
	By our choice of $c$ and $n$, we know $c \ge 1$ and $\log^{(2c)}n \ge \log^{(\log^*n-4)}n \ge 11$.
	Then the result follows directly from Corollary \ref{cor:winet}.
\end{proof}

\chapter{Conclusion}\label{sec:conclusion}

\section{Summary}\label{subsec:final_words}

We have shown in this thesis how to generate a Gray code, while reading
significantly fewer bits on average than previously known algorithms,
and how to efficiently generate a quasi-Gray code
with the same worst-case performance and improved space efficiency.
We give a tradeoff between space-efficiency, and the worst-case number of
bits written.  When the worst case number of bits is equal to the dimension
of the code, $d$, it is easy to generate all $2^d$ bit strings.
Our Recursive Partition Gray Code gives a $d$-dimensional Gray code,
writing at most a
single bit to generate each successive bit string, while reading no more than
$6\log d$ bits on average and with an optimal space efficiency of $1$.
As with all known algorithms that generate a Gray code, this requires
all $d$ bits are read in the worst case.

But when
the worst-case number of bits read decreases, our space-efficiency does also.
As an improvement on previous results, we present algorithms that are able to
retain space efficiency arbitrarily close to $1$ by reading more than a
constant number of bits on average.
Our \texttt{WineIncrement} algorithm gives an algorithm to generate a
quasi-Gray code with dimension $d = n+ (t+1) \log n$.
This algorithm reads on average $12\log^{(2c)}n + O(1)$ bits, while in the worst
case it reads only $(t+1)\log n + 1$ bits.  This is done with a space efficiency
of $1-O(n^{-t})$.  The quasi-Gray code changes in at most $2c+1$ bits to
reach any successive state in the code.

This trade-off highlights the initial problem which motivated this work: a
lower bound on the number of bits read in the worst case, for a
Gray code with space efficiency of $1$.
Fredman \cite{Fredman78} showed a lower bound of $\Omega(\log d)$ bits
read in the worst case, and improving this bound remains an open problem.

\begin{figure}
\label{fig:cube}
\begin{center}
\includegraphics{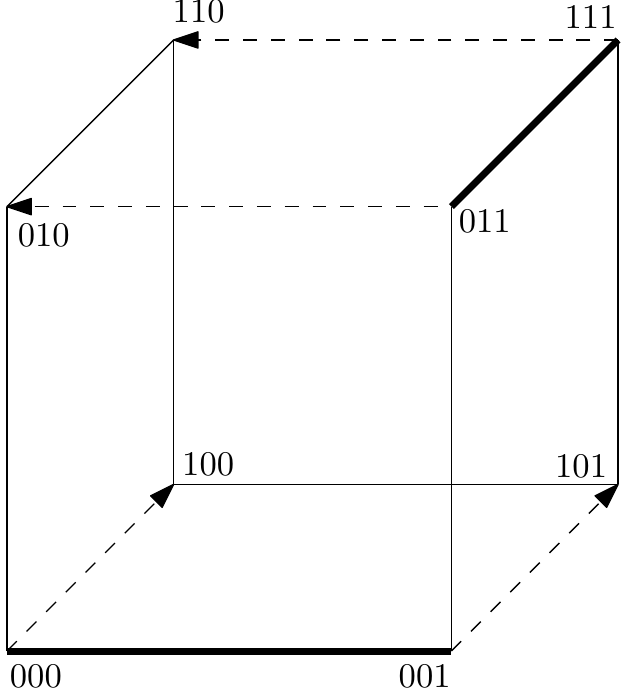}
\end{center}
\caption{Rules for traversing a hypercube of dimension $d=3$ that require
reading $d-1$ bits.  The rules are directed dashed lines, each of which
shows the bit that would be changed when the code's current bit string is
equal to the label of the rule's source vertex in the hypercube.
A dark line connects two states that differ only in the bit not read for
the two paired rules adjacent to it.}
\end{figure}

During our efforts to find a better lower bound,
we used extensively the idea of a Gray code
being equivalent to a Hamiltonian path (or cycle, for a cyclic code)
on a hypercube \cite{Leighton}.  We aimed to find an example of a Gray code
of dimension $d$, which read at most $d-1$ bits in the worst case,
or alternatively to show that such a code is not possible.
When traversing a hypercube, a \emph{rule} is a
directed edge between two points on the hypercube, $x$ and $y$, with the point
$x$ corresponding to a path in the DAT.
If every path in the DAT has length $d-1$, then it would be used
in exactly two situations: when the bit it did not read is either
a $0$ or a $1$.
In the hypercube, this means that any rule we add, from $x$ to $y$,
must be paired with a second parallel rule starting from a position one hop away
from $x$ and ending one hop away from $y$.
Thus all rules must come in pairs,
travelling parallel to each other, in the same direction, and adjacent to
a shared face on the hypercube.  Figure \ref{fig:cube} shows an example of
a hypercube with dimension three, and paired rules along its edges.
Note that only one rule may exist on any one edge of the hypercube.

This
model is weaker than the DAT model, as not all arrangements of rules on the
hypercube may be constructed as a DAT, however all DATs of height
$d-1$ can be converted to pairs of rules on the hypercube.
Thus, showing such a construction
is not possible on the hypercube would also show it is not possible for a DAT.
If paired rules, such as those in Figure \ref{fig:cube},
cannot be placed without overlapping
such that a Hamiltonian (cycle) path is formed on the hypercube, then it is also
not possible to construct a DAT that generates a (cyclic) Gray code of dimension
$d$ without reading $d$ bits in the worst case.
For small $d$, it quickly becomes obvious that such pairs of rules cannot be
used to construct a Hamiltonian cycle on the hypercube.  For larger dimensions,
we conjecture that it is never possible to construct a Hamiltonian cycle
using pairs of parallel face-adjacent rules on the hypercube.

\section{Future work}

The main open problem remains: finding a better lower bound for the worst-case
number of bits read by a Gray code with space efficiency of $1$, or finding
an example of a such Gray code that always reads less than $d$ bits to
generate the next bit string in the code.

When considering the average number of bits read, is it possible to
find a Gray code with space efficiency $1$ that requires reading
less bits than the Recursive Partition Gray Code?
As a Gray code, what properties does the RPGC have, such as
run length, balance of bit flips, and others discussed by Savage? \cite{Savage}

What other properties does
the Recursive Partition Gray Code have, such as 

Is it possible to find a quasi-Gray code with space efficiency $1$, or a
quasi-Gray code that reads fewer bits on average or in the worst case
than the \texttt{WineIncrement} algorithm?  Is there a lower bound on the space
efficiency in relation to the number of bits read on average?
Fredman \cite{Fredman78} showed a tradeoff between the number of bits written
and read in the worst case.  Can a tighter tradeoff be shown, between the
worst-case number of bits written, the worst-case number of bits read,
the average number of bits read, and/or the space efficiency of a quasi-Gray
code?

Our Recursive Partition Gray Code does provide a counter-example to any
efforts to show a best-case bound of more than $\Omega(\log d)$, and our hope
is that this work will contribute to a better understanding of
the problem, and eventually, a tighter lower bound in the case of
generating a space-optimal Gray code.

\cleardoublepage
\bibliographystyle{alpha}
\bibliography{graycodes}
\end{document}